\documentclass{lmcs}
\pdfoutput=1

\usepackage{lastpage}
\lmcsdoi{16}{1}{16}
\lmcsheading{}{\pageref{LastPage}}{}{}%
{Aug.~13,~2018}{Feb.~13,~2020}{}

\usepackage{mathdefs,relsize,amssymb}

\let\sc\scshape

\makeatletter
\newcommand\m@thsm@ller[2]{\mbox{\relscale{0.91}$\m@th#1#2$}}
\let\smaller\undefined%
\DeclareRobustCommand\smaller[1]{\relax\ifmmode{\mathpalette\m@thsm@ller{#1}}\else{\relscale{0.91}#1}\fi}
\makeatother

\DeclareRobustCommand*{\dom}{\qopname\relax o{dom}}
\DeclareRobustCommand*{\rng}{\qopname\relax o{rng}}
\newcommand*{\id}{\mathrm{id}}
\newcommand*{\Flat}{\mathrm{flat}}
\newcommand*{\sing}{\mathrm{sing}}
\newcommand*{\MSO}{\smaller{\mathrm{MSO}}}
\renewcommand*{\PSet}{\mathcal{P}}
\newcommand*{\emptyseq}{\langle\rangle}
\newcommand*{\?}{\kern .08em} % chktex 1
\newcommand*{\napprox}{\not\approx}

\keywords{infinite trees, tree algebras, regular languages, monads}

\begin{document}
\title{Regular Tree Algebras}
\author[A.~Blumensath]{Achim Blumensath}
\address{Masaryk University Brno, Czech Republic}
\email{blumens@fi.muni.cz}
\thanks{Work supported by the Czech Science Foundation, grant No.~GA17-01035S} % chktex 8

\begin{abstract}
We introduce a class of algebras that can be used as recognisers for regular tree languages.
We show that it is the only such class that forms a pseudo-variety and we prove the
existence of syntactic algebras. Finally, we give a more algebraic characterisation
of the algebras in our class.
\end{abstract}

\maketitle

\section{Introduction}   %%%%%%%%%%%%%%%%%%%%%%%%%%%%%%%%%%%%%%%%%%%%%%%%%%%%%%%

There are many different formalisms to study regular languages,
the most prominent ones being automata and logic.
In this paper we are interested in the algebraic approach to formal language theory,
in the context of infinite trees.
Such algebraic methods are particularly successful in deriving decidable
characterisations for various fragments of monadic second-order logic.
For instance, a theorem of Sch\"utzenberger~\cite{Schutzenberger:1965il}
states that a language of finite words is definable in first-order logic if,
and only if, its syntactic monoid is finite and aperiodic.
The latter condition is decidable as we can compute the syntactic monoid of a
regular language and check it for aperiodicity.

Besides a comprehensive algebraic theory for the usual word languages,
there also exist well-developed frameworks for languages of infinite words and -- to a lesser degree -- finite trees. % chktex 8
For languages of infinite trees, the combinatorics involved are much more challenging.
As a result, the existing theory is still fragmentary.
The first preliminary results were provided in~\cite{BojanczykId09,BojanczykIdSk13}, with one article considering languages of
\emph{regular} trees only, and one considering languages of \emph{thin} trees.
The first framework that could deal with arbitrary infinite trees
was provided by~\cite{Blumensath11c,Blumensath13a}.
Unfortunately, it turned out to be too complicated and technical to be very useful.

In this article we propose an alternative, much simpler approach,
and we develop it to a point where it is suitable for devising decision procedures.
Because of space considerations we defer the actual applications to a subsequent
article~\cite{BlumensathYY}.
Our first simplification concerns the notation. It turns out that much of the
notational overhead of the old framework can be avoided by adopting
the category-theoretical formalism of a monad and an Eilenberg-Moore algebra.
Our second contribution is in isolating a suitable class of algebras
as recognisers of regular languages.
While admittedly its definition is rather na\"{\i}ve and not as concrete
as one would like it to be,
our key insight -- and the main contribution of this paper -- is the fact that % chktex 8
the resulting class has all the required properties\?: it forms a pseudo-variety
and it has syntactic algebras. Furthermore, we prove that it is the only
class that does the trick (cf.~Corollary~\ref{Cor: uniqueness} below).

The overview of this article is as follows.
We start in Section~\ref{Sect: tree algebras} with setting up our algebraic framework.
In particular, we explain the notion of an Eilenberg--Moore algebra. % chktex 8
In Section~\ref{Sect: regular algebras} we isolate the property (`regularity')
we need for a tree algebra to recognise regular languages only, and we give a first
characterisation of when an algebra has this property.
While both the definition and our characterisation are rather abstract,
we show that the resulting class is the only possible one that satisfies all our
requirements\?:
we prove in Section~\ref{Sect: pseudo-varieties} that it is
the only class with the desired closure properties\?; and
in Section~\ref{Sect: syntactic algebras} we prove the existence of syntactic algebras,
a prerequisite for characterisation results.
We conclude in Section~\ref{Sect: deterministic}
with a second, more specific characterisation of regularity for tree algebras.

\paragraph{\bf Acknowledgements}
This paper owes much to unpublished work of and discussions with Boja\'n\-czyk and Klin
who gracefully allowed me to include their results. In particular the proof of
Theorem~\ref{Thm:syntactic congruence is one} is entirely due to them.
As it is rather hard to separate their contributions from my own,
I~have refrained from adding attributions to specific results.
Instead Boja\'n\-czyk and Klin should be considered co-authors in spirit,
even if they chose not to be listed as such.

\section{Tree algebras}%  %%%%%%%%%%%%%%%%%%%%%%%%%%%%%%%%%%%%%%%%%%%%%%%%%%%%%%%%%%%%%%%%
\label{Sect: tree algebras}

A convenient algebraic formalism for the various kinds of language theories
has turned out to be one based on the category-theoretical notions of
a \emph{monad} and an \emph{Eilenberg-Moore algebra}~\cite{Bojanczyk15}.
To make this article accessible to readers without a category-theoretical background
we refrain from using category-theoretical terminology where possible and
use elementary definitions instead. Readers familiar with category theory
should be able to translate our results into their language.

To prepare the reader for  our notion of a tree algebra, let us take a look at semigroups first.
Instead of using the usual binary product, we can see a semigroup as a set~$S$
equipped with a product $\pi : S^+ \to S$ of variable arity that multiplies
an arbitrary sequence of semigroup elements in one step.
Analogously, we will define a tree algebra as a set~$A$ together with a product
$\pi : \bbT A \to A$ that takes an $A$-labelled tree and returns a single element
of~$A$.
Let us make this idea precise.

First of all, we will not work with simple sets but with \emph{ranked sets,}
that is, sets where each element has an \emph{arity} or \emph{rank.}
Formally, we consider such a set as a sequence $A = {(A_n)}_{n<\omega}$
where $A_n$~is the subset of elements of arity~$n$.
A~function $f : A \to B$ of such sets is then a family $f = {(f_n)}_{n<\omega}$
of functions $f_n : A_n \to B_n$.
We will frequently identify a ranked set $A = {(A_n)}_n$ with its disjoint union
$A = \bigcupdot_n A_n$. A function ${(f_n)}_n : {(A_n)}_n \to {(B_n)}_n$ then corresponds
to a rank preserving function $f : A \to B$.
In the rest of this article all sets will be assumed to be ranked,
if not explicitly stated otherwise, and all functions will be rank preserving.

Now let $A$~be a ranked set.
An \emph{$A$-labelled} tree~$t$ is a (finite or infinite) tree where every
vertex is labelled by an element from~$A$ in such a way that the arity of a label
matches the number of successors.
We set $\bbT A = {(\bbT_n A)}_{n<\omega}$ where $\bbT_n A$~is the set of all
$(A \cup \{x_0,\dots,x_{n-1}\})$-labelled trees~$t$ where the additional labels~$x_i$
are called \emph{variables.} These are considered as having arity~$0$
and we require that
\begin{itemize}
\item each variable~$x_i$ occurs at most once in the tree~$t$ and
\item the root is not labelled by any variable.
\end{itemize}
(We will always assume that $x_i \notin A$.)
Note that a tree~$t$ containing the variables, say, $x_0$,~$x_3$, and~$x_7$,
can be regarded as an element of $\bbT_8 A$, of $\bbT_9 A$, and so on.
According to our convention these elements are considered to be different.
For a tree~$t$, we denote its set of vertices by $\dom(t)$, and we write $t(v)$
for the label of a vertex $v \in \dom(t)$.
We say that two trees \emph{have the same shape} if there become equal when
we remove all non-variable labels.
\begin{rem}
There is some freedom in choosing how to define~$\bbT A$.
Instead of requiring that every variable occurs at most once,
we could allow each occurring several, even infinitely many times.
We also could allow the use of infinitely many different variables by adding
elements of arity~$\omega$. Finally, we could require that every variable
appears at least once.
For most of our results, these details do not matter.
Hence, the precise definition is more of a matter of taste.
But let us mention that some results in Section~\ref{Sect: deterministic}
fail if we allow multiple occurrences of the same variable.
In particular, this is the case for Proposition~\ref{Prop: reduction for arities}.
\end{rem}

To write down trees concisely we use the usual term notation.
For instance, $a(x_3,c)$ denotes the tree where the root is labelled by~$a$
and its two successors by $x_3$~and~$c$, respectively.
Another useful piece of notation is the following one.
Given a (rank-preserving) function $f : A \to B$ we denote by $\bbT f : \bbT A \to \bbT B$
the function that applies~$f$ to every (non-variable) label of the input tree.

Now we can define a \emph{tree algebra} $\frakA = \langle A,\pi\rangle$
as a ranked set~$A$ together with a product $\pi : \bbT A \to A$ that satisfies
certain \emph{associativity laws.}
Before stating these laws formally let us again take a look at semigroups.
For a function $\pi : S^+ \to S$ to be the product associated with a semigroup
it has to satisfy two conditions. First of all, we require that
the product is the identity on singletons, that is,
\begin{align*}
  \pi(\langle a\rangle) = a\,,
  \quad\text{for every } a \in S\,.
\end{align*}
Let us call this the \emph{unit law.}
Secondly, if we factorise a product in different ways, we always get the same result.
That means, for a sequence of sequences
$w = \langle w_0,\dots,w_{m-1}\rangle \in {(S^+)}^+$, we require that
\begin{align*}
  \pi\bigl(\pi(w_0),\dots,\pi(w_{m-1})\bigr) = \pi(w_0\dots w_{m-1})\,.
\end{align*}
Writing $\pi^+ : {(S^+)}^+ \to S^+$ for the function that multiplies each component
of the given sequence and
$\Flat : {(S^+)}^+ \to S^+$ for the concatenation function,
we can write this equation in the compact form

\noindent
\begin{minipage}[t]{0.55\textwidth}
\vspace*{0pt}%
\abovedisplayskip=1ex
\begin{align*}
  \pi \circ \pi^+ = \pi \circ \Flat\,.
\end{align*}
\medskip\noindent
This is the \emph{associative law.}
\end{minipage}%
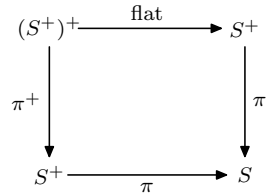
\begin{minipage}[t]{0.3\textwidth}
\vspace*{0pt}%
\centering
\includegraphics{Regular-submitted-1.mps}
%\begin{mpfig}
%  u := 1.3cm;
%
%  z0 = (0,0);
%  z1 = (2u,0);
%  z2 = (0,1.5u);
%  z3 = (2u,1.5u);
%
%  pickup pencircle scaled 0.6pt;
%
%  drawarrow anchor(z0,z1,7pt) -- anchor(z1,z0,7pt);
%  drawarrow anchor(z2,z0,7pt) -- anchor(z0,z2,7pt);
%  drawarrow anchor(z2,z3,11pt) -- anchor(z3,z2,9pt);
%  drawarrow anchor(z3,z1,7pt) -- anchor(z1,z3,7pt);
%
%  label (btex $S^+$     etex, z0);
%  label (btex $S$       etex, z1);
%  label (btex $(S^+)^+$ etex, z2);
%  label (btex $S^+$     etex, z3);
%
%  label bot (btex $\pi$   etex, 1/2[z0,z1]);
%  label lft (btex $\pi^+$ etex, 1/2[z2,z0]);
%  label top (btex $\Flat$ etex, 1/2[z2,z3]);
%  label rt  (btex $\pi$   etex, 1/2[z3,z1]);
%\end{mpfig}
\end{minipage}

\medskip
Introducing the corresponding auxiliary functions for trees,
we can write similar laws for a tree algebra $\frakA = \langle A,\pi\rangle$\?:

\noindent
\begin{minipage}[t]{0.55\textwidth}
\vspace*{0pt}%
\abovedisplayskip=1ex
\begin{align*}
  \pi \circ \sing &= \id \\[1ex]
\prefixtext{and}
  \pi \circ \bbT\pi &= \pi \circ \Flat\,.
\end{align*}
\end{minipage}%
\begin{minipage}[t]{0.3\textwidth}
\vspace*{0pt}%
\centering
\includegraphics{Regular-submitted-2.mps}
%\begin{mpfig}
%  u := 1.3cm;
%
%  z0 = (0,0);
%  z1 = (2u,0);
%  z2 = (0,1.5u);
%  z3 = (2u,1.5u);
%
%  pickup pencircle scaled 0.6pt;
%
%  drawarrow anchor(z0,z1,7pt) -- anchor(z1,z0,7pt);
%  drawarrow anchor(z2,z0,7pt) -- anchor(z0,z2,7pt);
%  drawarrow anchor(z2,z3,11pt) -- anchor(z3,z2,9pt);
%  drawarrow anchor(z3,z1,7pt) -- anchor(z1,z3,7pt);
%
%  label (btex $\bbT A$     etex, z0);
%  label (btex $A$          etex, z1);
%  label (btex $\bbT\bbT A$ etex, z2);
%  label (btex $\bbT A$     etex, z3);
%
%  label bot (btex $\pi$     etex, 1/2[z0,z1]);
%  label lft (btex $\bbT\pi$ etex, 1/2[z2,z0]);
%  label top (btex $\Flat$   etex, 1/2[z2,z3]);
%  label rt  (btex $\pi$     etex, 1/2[z3,z1]);
%\end{mpfig}
\end{minipage}

\begin{figure}
\centering
\includegraphics{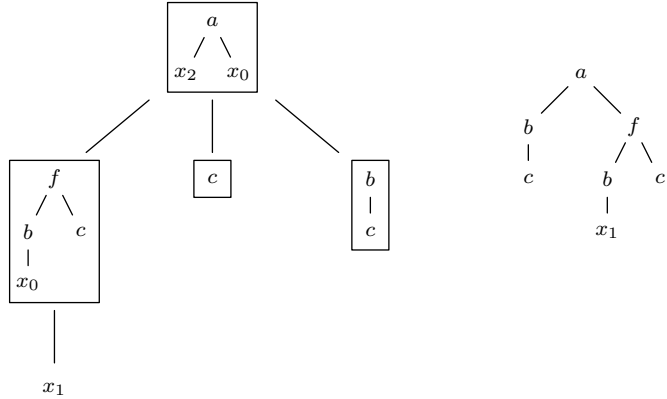}
\caption{The flattening operation}\label{Fig: flat}
\end{figure}

\medskip\noindent
Here, the \emph{singleton function} $\sing : A \to \bbT A$ maps a label $a \in A_n$
to the \emph{singleton tree} $a(x_0,\dots,x_{n-1}) \in \bbT_{n}A$, and the
\emph{flattening function} $\Flat : \bbT\bbT A \to \bbT A$ takes a tree~$t$
whose vertices~$v$ are labelled by trees~$t(v)$ from~$\bbT A$ and returns
the tree obtained by simultaneously substituting in~$t(v)$ each variable~$x_i$
by the tree associated with the corresponding successor of~$v$.
In more detail, we compute $\Flat(t)$ as follows.
We start with the disjoint union of all trees $t(v)$, for $v \in \dom(t)$.
We then remove every leaf of (the copy of) $t(v)$ that is labelled by a variable~$x_i$,
and replace it with an edge to the root of the corresponding copy of~$t(u_i)$,
where $u_i$~is the $(i+1)$-th successor of~$v$.
Of the resulting forest, we take the connected component containing the root~$t(\emptyseq)$.
This is the value of the flattening $\Flat(t)$.
For instance, in Figure~\ref{Fig: flat} the tree on the left evaluates
to the tree on the right.

\smallskip
To summarise let us give the formal definition.
\begin{defi}
(a) A \emph{tree algebra} is a pair $\frakA = \langle A,\pi\rangle$
consisting of a ranked set~$A$ and a \emph{product} $\pi : \bbT A \to A$ which satisfy
\begin{align*}
  \pi \circ \sing = \id
  \qtextq{and}
  \pi \circ \bbT\pi = \pi \circ \Flat\,.
\end{align*}

(b)
A \emph{morphism} of tree algebras is a function $f : A \to B$ between
their domains commuting with the respective products\?:

\noindent
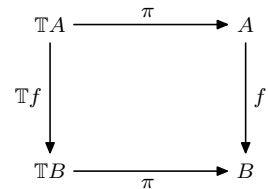
\begin{minipage}[t]{0.55\textwidth}
\vspace*{-1ex}%
\begin{align*}
  f \circ \pi = \pi \circ \bbT f\,.
\end{align*}
\end{minipage}%
\begin{minipage}[t]{0.45\textwidth}
\vspace*{0pt}%
\centering
\includegraphics{Regular-submitted-4.mps}
%\begin{mpfig}
%  u := 1.3cm;
%
%  z0 = (0,0);
%  z1 = (2u,0);
%  z2 = (0,1.5u);
%  z3 = (2u,1.5u);
%
%  pickup pencircle scaled 0.6pt;
%
%  drawarrow anchor(z0,z1,9pt) -- anchor(z1,z0,7pt);
%  drawarrow anchor(z2,z0,7pt) -- anchor(z0,z2,7pt);
%  drawarrow anchor(z2,z3,9pt) -- anchor(z3,z2,7pt);
%  drawarrow anchor(z3,z1,7pt) -- anchor(z1,z3,7pt);
%
%  label (btex $\bbT B$ etex, z0);
%  label (btex $B$      etex, z1);
%  label (btex $\bbT A$ etex, z2);
%  label (btex $A$      etex, z3);
%
%  label bot (btex $\pi$    etex, 1/2[z0,z1]);
%  label lft (btex $\bbT f$ etex, 1/2[z2,z0]);
%  label top (btex $\pi$    etex, 1/2[z2,z3]);
%  label rt  (btex $f$      etex, 1/2[z3,z1]);
%\end{mpfig}
\end{minipage}
\end{defi}

\noindent
An important example of a tree algebra is a free one.
Given a ranked set~$\Sigma$, the \emph{free tree algebra} over~$\Sigma$ is
$\langle\bbT\Sigma, \Flat\rangle$.
(The facts that this is indeed a tree algebra and that it has the
desired universal property
follow from a general category-theoretical result on monads\?;
see e.g.\ Proposition~4.1.4 of~\cite{Borceux94b}.)

\smallskip
Let us next explain how to use a tree algebra to recognise tree languages.
For the purpose of this article, a \emph{tree language} is a subset
$L \subseteq \bbT_m\Sigma$ where $\Sigma$~is a finite ranked set and the arity $m < \omega$
is fixed. A tree language $L \subseteq \bbT_m\Sigma$ is called \emph{regular}
if it is recognised by a nondeterministic tree automaton with the parity
condition, or it is definable by a formula of monadic second-order logic,
see~\cite{Thomas97}.  (For $m>0$, the automaton or formula treats the
variables as distinguished letters in the leaves.)
Such a language is \emph{recognised} by a morphism $\varphi : \bbT\Sigma \to \frakA$
if there is a subset $P \subseteq A_m$ such that $L = \varphi^{-1}[P]$.
In this case, we also say that the algebra~$\frakA$ recognises~$L$.

As an example, let us construct a tree algebra recognising the set of all trees
$t \in \bbT_0\{a,b\}$ that contain the label~$a$ at least once.
For every (part of an) input tree, we have to remember one bit of information\?:
whether or not it contains the label~$a$.
This suggests to have two elements, say $0_m$~and~$1_m$, for each arity~$m$.
When constructing arbitrary products of such elements, we obtain additional elements
that are of the form $c(x_i,x_j,\dots,x_k)$ where $c \in \{0_m,1_m\}$ and
$i < j < \cdots < k < m$.
(As it is important to know which variables appear in a term, we cannot simply
identify these with $0_m$~and~$1_m$.)
Thus, the domain~$A_m$ for arity~$m$ of our algebra~$\frakA$ will consist of all
these elements and the recognising morphism maps a term $t \in \bbT_m\{a,b\}$
to the element $c(x_i,x_j,\dots,x_k)$ where $c$~specifies whether or not $t$~contains
the label~$a$ and $x_i,x_j,\dots,x_k$ are the variables that actually appear in~$t$.

\section{Regular tree algebras}%   %%%%%%%%%%%%%%%%%%%%%%%%%%%%%%%%%%%%%%%%%%%%%%%%%%%%%%%%
\label{Sect: regular algebras}

The goal of this paper is to find a class~$\calC$ of tree algebras that
\emph{characterises} the class of regular tree languages in the sense that
a tree language is regular if, and only if, it is recognised by some algebra
from~$\calC$.
One obvious condition we have to impose on such a class is that all algebras
$\frakA \in \calC$ are \emph{finitary,} which means that
\begin{itemize}
\item $\frakA$ is finitely generated (i.e., there is a finite set $C \subseteq A$
  such that every element $a \in A$ can be written as a product of some tree
  in~$\bbT C$) and
\item for every $n < \omega$, there are only finitely many elements of arity~$n$.
\end{itemize}
Unfortunately, this in itself is not enough. There are examples of finitary tree
algebras that recognise non-regular languages~\cite{BojanczykKlin17}.

A na\"{\i}ve way to obtain the desired class of algebras
is to take the class of all tree algebras that
only recognise regular languages.
This is obviously the largest class that will do (if~any exists at all).
The problem with this definition is that it is not very enlightening\?:
we have no idea of what these algebras look like.
We will nevertheless adopt this na\"{\i}ve approach for its simplicity.
A~more satisfying, but also much more complicated, alternative definition
will be provided in Section~\ref{Sect: deterministic} below.

By looking at what it means to only recognise regular languages,
we arrive at the following definition.
\begin{defi}
A tree algebra $\frakA = \langle A,\pi\rangle$ is \emph{regular} if it is finitary
and there exists a finite set $C \subseteq A$ of generators such that, for every
element $a \in A$, the preimage
\begin{align*}
  \pi^{-1}(a) \cap \bbT C \quad\text{is a regular language.}
\end{align*}
\end{defi}
Note that an alternative way to write the set in the above definition is
${(\pi \restriction \bbT C)}^{-1}(a)$,
where $\pi \restriction X$ denotes the restriction of~$\pi$ to the set~$X$.
This will come in handy in several of the proofs below.

Before showing that our definition has the desired effect, let us mention that
it does not depend on the choice of the set~$C$ of generators.
\begin{lem}\label{Lem: regularity does not depend on the set of generators}
Let $\frakA$~be a regular tree algebra and $D \subseteq A$ a finite set.
Then
\begin{align*}
  \pi^{-1}(a) \cap \bbT D \text{ is regular,}\quad
  \text{for all } a \in A\,.
\end{align*}
\end{lem}
\begin{proof}
For each $d \in D$, we fix some term $t \in \bbT C$ with $\pi(t) = d$.
This defines a function $s_0 : D \to \bbT C$ such that $\pi \circ s_0 = \id$.
We can extend~$s_0$ to a morphism $s : \bbT D \to \bbT C$ by setting
\begin{align*}
  s(t) = \Flat(\bbT s_0(t))\,.
\end{align*}
For $t \in \bbT D$, it follows that
\begin{align*}
  (\pi \circ s)(t)
   = (\pi \circ \Flat \circ \bbT s_0)(t)
  &= (\pi \circ \bbT\pi \circ \bbT s_0)(t) \\
  &= (\pi \circ \bbT(\pi \circ s_0))(t)
   = (\pi \circ \bbT\id)(t)
   = \pi(t)\,,
\end{align*}
which implies that $\pi \restriction \bbT D = (\pi \restriction \bbT C) \circ s$.
For $a \in A$, we therefore have
\begin{align*}
  {(\pi \restriction \bbT D)}^{-1}(a)
  = s^{-1}[{(\pi \restriction \bbT C)}^{-1}(a)]\,.
\end{align*}
By assumption the set ${(\pi \restriction \bbT C)}^{-1}(a) = \pi^{-1}(a) \cap \bbT C$ is regular.
As regular languages are closed under inverse morphisms
(see Lemma~\ref{Lem: closure under inverse morphisms}),
so is ${(\pi \restriction \bbT D)}^{-1}(a)$.
\end{proof}

Let us now establish the rather obvious fact that our definition does what it is supposed to.
\begin{thm}\label{Thm:regular algebra iff recognises regular languages}
A finitary tree algebra is regular if, and only if,
all languages recognised by it are regular.
\end{thm}
\begin{proof}
$(\Leftarrow)$
Suppose that $\frakA$~is not regular.
Then there exists a finite set $C \subseteq A$ and an element $a \in A$
such that the preimage $L := \pi^{-1}(a) \cap \bbT C$ is not regular.
Consequently,
the restriction $\pi \restriction \bbT C : \bbT C \to \frakA$
of the product is a morphism that recognises a non-regular language~$L$.

$(\Rightarrow)$
Suppose that $\frakA$~is regular and let
$\varphi : \bbT\Sigma \to \frakA$ be a morphism recognising the language
$L := \varphi^{-1}[P]$ with $P \subseteq A_m$.
Set $C := (\varphi \circ \sing)[\Sigma]$.
By Lemma~\ref{Lem: regularity does not depend on the set of generators},
each preimage
\begin{align*}
  K_a := \pi^{-1}(a) \cap \bbT C\,, \quad\text{for } a \in A\,,
\end{align*}
is regular.
Hence, so is the (finite) union $K := \bigcup_{a \in P} K_a$.
Let $i : C \to A$ be the inclusion map,
$\pi_0 := \pi \restriction \bbT C : \bbT C \to A$ the restriction of the product,
and set $\varphi_0 := \varphi \circ \sing : \Sigma \to C$.
It follows that
\begin{align*}
  \varphi \circ \sing
  = i \circ \varphi_0
  = \pi_0 \circ \sing \circ \varphi_0
  = \pi_0 \circ \bbT\varphi_0 \circ \sing\,.
\end{align*}
{\centering
\includegraphics{Regular-submitted-5.mps}
\par}

\medskip
Since $\bbT\Sigma$ is generated by the range of $\sing$,
this implies that $\varphi = \pi_0 \circ \bbT\varphi_0$.
Hence,
\begin{align*}
  L = \varphi^{-1}[P]
  = {(\pi_0 \circ \bbT\varphi_0)}^{-1}[P]
  = {(\bbT\varphi_0)}^{-1}[\pi_0^{-1}[P]]
  = {(\bbT\varphi_0)}^{-1}[K]\,.
\end{align*}
As regular languages are closed under inverse projections
(see Lemma~\ref{Lem: closure under inverse morphisms}),
it therefore follows that $L$~is regular.
\end{proof}

Conversely one can prove that every regular language is recognised by some
regular tree algebra.
\begin{thm}\label{Thm:regular algebras characterise the regular languages}
A tree language is regular if, and only if, it is recognised by a regular
tree algebra.
\end{thm}
One direction follows immediately from
Theorem~\ref{Thm:regular algebra iff recognises regular languages}.
For the other one, we have to construct a regular tree algebra
recognising a given regular language~$L$.

We start by fixing notation and collecting a few basic definitions
(more details can be found in, e.g.,~\cite{Niesser02}).
We work with non-deterministic parity automata of the form
$\calA = \langle Q,\Sigma,\Delta,q_0,\Omega\rangle$,
where $Q$~is the (unranked) set of states, $\Sigma$~the (ranked) input alphabet,
$q_0$~the initial state, $\Omega : Q \to \omega$ a priority function,
and $\Delta \subseteq Q \times \Sigma \times Q^*$ the transition relation.
Each transition $\langle q,a,p_0,\dots,p_{n-1}\rangle$ consists of the
current state~$q$, the current letter~$a$, and states $p_0,\dots,p_{n-1}$
for the successors. For leaves, the letter~$a$ has arity~$0$ and the transition
simply takes the form $\langle q,a\rangle$.

A \emph{partial run} of~$\calA$ on some input tree $t \in \bbT\Sigma$ is
a labelling $\varrho : \dom(t) \to Q$ of the tree such that
\begin{itemize}
\item there are arbitrary states at vertices carrying a variable~$x_i$,
\item the labelling respects the transition relation~$\Delta$ at all other vertices, and
\item every infinite branch satisfies the parity condition.
\end{itemize}
The \emph{profile} of a partial run~$\varrho$ is the tuple
\begin{align*}
  \langle q,k_0,p_0\dots,k_{m-1},p_{m-1}\rangle\,,
\end{align*}
where $q$~is the state at the root of~$t$,
$p_i$~the state at the vertex carrying the variable~$x_i$,
and $k_i$~the minimal priority seen along the path from the root to this vertex.
If there is no vertex labelled~$x_i$, we set $k_i := \bot$ and $p_i := \bot$,
for some special bottom symbol~$\bot$.

We aim to construct a tree algebra where the elements encode sets of possible profiles,
i.e., sets of possible behaviours of~$\calA$ on a given input tree.
To simplify the definition and accommodate the material in
Section~\ref{Sect: deterministic} below,
we will construct an algebra that is slightly larger than necessary\?:
instead of using only the usual profiles of~$\calA$,
we will work with \emph{partial} ones, i.e., profiles where we only specify
data for some of the variables.
Formally, this can be done by labelling the paths to the variables
by elements of a suitable $\omega$-semigroup (for the definition of an $\omega$-semigroup,
see, e.g.,~\cite[Section 4.1]{PerrinPin04}).
The construction is performed in three steps.
(We keep the presentation rather informal. More details can be found in~\cite{BlumensathXX}.)

(i)
We denote by $\frakS_\calA = \langle S,S_\omega\rangle$ the (partial) $\omega$-semigroup where
\begin{itemize}
\item $S := Q \times D \times Q$ contains all triples of the form
  $\langle p,k,q\rangle$ for states $p,q \in Q$ and a priority~$k$
  ($D$~is the set of priorities used by~$\calA$) and
\item $S_\omega := Q$ contains the states of~$\calA$.
\end{itemize}
A triple $\langle p,k,q\rangle \in S$ encodes a finite path of a run
that starts in state~$p$, ends in state~$q$, and has minimal priority~$k$.
A state $p \in S_\omega$ encodes an infinite branch that starts
in state~$p$ and satisfies the parity condition.

The product is defined naturally\?: if we multiply two triples
$\langle p,k,q\rangle$ and $\langle p',k',q'\rangle$ with matching states $p' = q$,
the result is $\langle p, \min(k,k'), q'\rangle$.
If $p' \neq q$, the product remains undefined.
Similarly, the product of $\langle p,k,q\rangle$ and $p' \in S_\omega$
evaluates to $p \in S_\omega$, provided that $q = p'$.
Otherwise, it is again undefined.
Finally, an infinite product of a sequence $\langle p_i,k_i,q_i\rangle_{i<\omega}$
produces the state~$p_0$, provided that
$q_i = p_{i+1}$ for all~$i$ and the parity condition
\begin{align*}
  \liminf_{i \to \infty} \Omega(p_i) \text{ is even}
\end{align*}
is satisfied.

(ii) Next we turn $\frakS_\calA$ into a tree algebra
where the elements of arity~$m$ are of the form $a(x_i)$ or~$b$ with
$a \in S$, $b \in S_\omega$, and $i < m$.
The product is induced by the $\omega$-semigroup product\?:
given a tree~$t$ labelled by elements of this form, we construct a branch
by starting at the root and proceeding downwards as follows.
If the label of the current vertex is $a(x_i)$, we continue with the $(i+1)$-th
successor. If it is of the form $b \in S_\omega$, we stop.
This process yields a sequence of elements of the $\omega$-semigroup,
which we can multiply to a new element~$c$. If the chosen branch ends in a
variable~$x_k$, we return $c(x_k)$, otherwise we simply return~$c$.
We leave the product of~$t$ undefined,
if the product of the $\omega$-semigroup elements is not defined.

(iii) Finally, we formally close the tree algebra constructed in~(ii) first under conjunctions,
and then under disjunctions, that is, the new elements are formal expressions of the from
$\Lor_i\Land_k a_{ik}$ where the elements~$a_{ik}$ all have the same arity.
We define a product of such elements by requiring that disjunction and conjunctions
commute with the product operation. If a product of basic elements is undefined,
we treat it as an empty conjunction.

\smallskip
Let $\frakA$~be the tree algebra constructed in~(iii).
(It is straightforward, but rather tedious, to check that
$\frakA$~is indeed a tree algebra, i.e., that the product is associative.
The interested reader can find a full proof in~\cite{BlumensathXX}.)

To show that $\frakA$~recognises $L(\calA)$,
note that conjunctions of $\omega$-semigroup elements can be used to encode partial profiles
of~$\calA$ and disjunctions of such conjunctions sets of partial profiles.
Hence, we consider the morphism $\varphi : \bbT\Sigma \to \frakA$ that maps a tree~$t$
to the disjunction $\Lor_\varrho \tilde\varrho$
where $\varrho$~ranges over all partial runs of~$\calA$ on~$t$
and $\tilde\varrho$~is an element encoding the run~$\varrho$ defined as follows.
Let $q$~be the state at the root,
${(v_i)}_i$~an enumeration of all vertices of~$t$ with a variable,
${(p_i)}_i$~the corresponding sequence of states, $x_{m(i)}$ the variable at~$v_i$,
and let $k_i$~be the minimal priority on the path from the root to~$v_i$.
We set
\begin{align*}
  \tilde\varrho := q \land \mathop{\smash{\Land_i}} {\langle q,k_i,p_i\rangle(x_{m(i)})\,.}
\end{align*}
It follows that $\calA$~accepts a tree~$t$ if, and only if,
\begin{align*}
  \varphi(t) \geq q_0 \land \mathop{\smash{\Land_i}} {\langle q_0,k_i,p_i\rangle(x_{m(i)})\,,}
\end{align*}
for some $p_i,k_i,m(i)$ such that, when starting in state~$p_i$,
the automaton~$\calA$ accepts the singleton tree with label~$x_{m(i)}$.
(The ordering~$\geq$ here is the one induced by the conjunctions and disjunctions.
In case of the above formula it simply means that $\varphi(t)$ is a disjunction
where one of the terms is a conjunction that contains the right-hand side as a subconjunction.)
Consequently, we can find a set $P \subseteq A$ such that
\begin{align*}
  L(\calA) = \varphi^{-1}[P]\,,
\end{align*}
as desired.

Finally, let $\frakA_0 \subseteq \frakA$ be the image of~$\varphi$.
We claim that~$\frakA_0$ is the desired regular tree algebra.
We already have a morphism $\varphi : \bbT\Sigma \to \frakA_0$ recognising~$L$.
Hence, it remains to show that $\frakA_0$~is regular.
Clearly, $\frakA_0$~is generated by the finite set $C := (\varphi \circ \sing)[\Sigma]$.
Consider an element $\Lor_i \Land_k a_{ik}$.
To check that a product $\pi(t)$ with $t \in \bbT C$ evaluates to this value we have to select,
for every conjunction $\Land_k a_{ik}$ and every vertex~$v$ of~$t$,
some term of the disjunction $t(v)$.
Then we have to multiply the corresponding $\omega$-semigroup elements along every branch of~$t$
and check that the result is equal to the corresponding element~$a_{ik}$.
This process can clearly be performed by a tree automaton.

\section{Closure properties}%   %%%%%%%%%%%%%%%%%%%%%%%%%%%%%%%%%%%%%%%%%%%%%%%%%%%%%%%%%%%
\label{Sect: pseudo-varieties}

So far, we have done nothing deep. The interesting realisation is
that our na\"{\i}ve definition is actually sufficient for applications\?:
the class of regular algebras has all the desired closure properties
and it allows the computation of syntactic algebras.
We start by taking a look at the closure properties. Syntactic algebras are
the topic of Section~\ref{Sect: syntactic algebras}.

Recall that a \emph{variety} is a class~$\calC$ of algebras that is closed under
the operations of taking\?: (i)~$\sfH$~homomorphic images (i.e., quotients),
(ii)~$\sfS$~subalgebras, and (iii)~$\sfP$~arbitrary products.
Equivalently, this can be written as the equation $\calC = \sfH\sfS\sfP(\calC)$.
It follows from the axioms that every variety is also closed under directed colimits
(see, e.g., Remark 3.6\,(6) of~\cite{AdamekRosicky94}).
Furthermore, the famous Variety Theorem of Birkhoff states that varieties are
exactly those classes of algebras that can be defined by systems of equations
(see, e.g., Theorem 3.9 of~\cite{AdamekRosicky94}).

If we are interested in classes of finite algebras only,
one has to adapt these definitions slightly.
Since the product operation~$\sfP$ can produce infinite algebras, we replace it
by the operation~$\sfP_\omega$ of taking finite products only.
This leads to the definition of a \emph{pseudo-variety,} which is a class~$\calC$
satisfying $\calC = \sfH\sfS\sfP_\omega(\calC)$.
For classes of finite algebras, closure under directed colimits is trivial.
There is also a variant of the Birkhoff Variety Theorem by Reiterman~\cite{Reiterman82}
that characterises pseudo-varieties as exactly those classes that can be defined by a system of
\emph{profinite} equations.

In our setting with infinitely many sorts, we are interested in classes of
\emph{finitary} algebras, and we are again forced to slightly modify the definitions.
The problem is that subalgebras and finite products of finitary algebras
are not necessarily finitely generated (we will provide counterexamples below).
Therefore we replace~$\sfS$ by the operation~$\sfS_\omega$ of taking
finitely-generated subalgebras only and
we require closure under $\sfH\sfS_\omega\sfP_\omega$.
As closure under directed colimits is not automatic anymore we also have to
add it as an extra requirement.
In fact a slightly weaker condition suffices\?: closure under \emph{rank-limits.}
We say that a tree algebra~$\frakA$ is the rank-limit of a sequence
${(\frakB_n)}_{n<\omega}$ of tree algebras if, for every $m < \omega$,
the algebras $\frakA$~and~$\frakB_n$, for $n \geq m$, are isomorphic
if we restrict them to elements of arity at most~$m$.
Note that closure under rank-limits is a rather natural condition.
For instance, it is satisfied by every class axiomatised by a set of equations.
One can show that, for classes of finitary algebras that are closed under quotients,
closure under rank-limits and under directed colimits are equivalent.
\begin{defi}
A \emph{pseudo-variety} of tree algebras is a class~$\calC$ of finitary
tree algebras that is closed under (i)~quotients,
(ii)~finitely generated subalgebras of finite products, and (iii)~rank-limits.
\end{defi}

We start by showing that the regular tree algebras form a pseudo-variety. At
the moment it is open whether there is an analogue to the Theorem of Reiterman
in our setting. There is a general result by Milius and Urbat~\cite{MiliusUrbat19}
which provides variety theorems for many classes, including the class of regular tree algebras.
But it uses a rather abstract notion of an equation and it remains to be worked out
how exactly such equations look like in our case.

Before continuing, let us introduce a bit of notation concerning rank-limits.
First, for a ranked set~$A$ and an arity~$k$, we set
$A_{<k} := A_0 \cup\dots\cup A_{k-1}$. We consider $A_{<k}$ as a ranked set
that has no elements of arity~$k$ or higher.
For our functor~$\bbT$ we similarly set
$\bbT_{<k}A := {(\bbT A_{<k})}_{<k}$.
Finally, for a tree algebra $\frakA = \langle A,\pi\rangle$ we denote by
$\frakA|_{<k}$ the algebra with domain $A_{<k}$ and product
$\pi \restriction \bbT_{<k}A : \bbT_{<k}A \to A_{<k}$.
Note that $\frakA|_{<k}$ is not a tree algebra (an algebra for the functor~$\bbT$)
as the product is not of the right form.
Instead it is an algebra for the functor~$\bbT_{<k}$ (a $\bbT_{<k}$-algebra
is defined by the same two laws as a tree algebra, except that we replace the
functor~$\bbT$ by~$\bbT_{<k}$ throughout).
With this notation we can say that $\frakA$~is a rank-limit of ${(\frakB^n)}_n$ if
\begin{align*}
  \frakA|_{<k} \cong \frakB^n_{<k}\,, \quad\text{for } k \leq n \leq \omega\,,
\end{align*}
where the isomorphism is understood as a $\bbT_{<k}$-algebra isomorphism.

\begin{thm}\label{Thm:regular algebras are pseudo-variety}
The class of regular tree algebras forms a pseudo-variety.
\end{thm}
\begin{proof}
The proof is straightforward.
We have to show that the class of regular tree algebras is closed under
\begin{enumerate}[label={\normalfont(\alph*)}]
\item finitely generated subalgebras of finite products,
\item rank-limits,
\item quotients.
\end{enumerate}

(a) The empty product has exactly one element~$1_m$ for each arity~$m$.
Given a subalgebra~$\frakA$ generated by some finite set~$C$ and some element $1_m \in A$,
we have
\begin{align*}
  \pi^{-1}(1_m) \cap \bbT C = \bbT_m C\,,
\end{align*}
which is regular.

Hence, it remains to consider a finitely generated subalgebra~$\frakA$ of a non-empty,
finite product $\prod_{i < n} \frakB^i$.
Let $C \subseteq A$ and $D^i \subseteq B^i$ be finite sets of generators.
Increasing the~$D^i$ if necessary, we may assume that $C \subseteq \prod_i D^i$.
Let $p_i : \prod_i B^i \to B^i$ be the projections.
For $t \in \bbT\prod_i D^i$ and $\bar a = {(a_i)}_i \in A \subseteq \prod_i B^i$, we have
\begin{align*}
  \pi(t) = \bar a
  \quad\iff\quad
  \pi(\bbT p_i(t)) = a_i \quad\text{for all } i\,.
\end{align*}
As the~$\frakB^i$ are regular, it follows that
\begin{align*}
  \pi^{-1}(\bar a) \cap \prod_i\bbT D^i
  = \bigcap_i {{(\bbT p_i)}^{-1}\bigl(\pi^{-1}(a_i) \cap \bbT D^i\bigr)}
\end{align*}
is regular. Since regular languages are closed under intersection,
the preimage
\begin{align*}
  \pi^{-1}(\bar a) \cap \bbT C
  = \pi^{-1}(\bar a) \cap \prod_i \bbT D^i \cap \bbT C
\end{align*}
is also regular.

(b)
Let ${(\frakB^n)}_n$ be a sequence of regular tree algebras with rank-limit~$\frakA$.
To show that $\frakA$~is regular, let $C \subseteq A$ be a finite set of generators
and $a \in A$.
Fix a number $k < \omega$, such that
\begin{align*}
  C \cup \{a\} \subseteq A_{<k} = B^n_{<k}\,, \quad\text{for } n \geq k\,.
\end{align*}
As $\frakB^k$~is regular, the preimage
$\pi^{-1}(a) \cap \bbT C$ is regular.

(c) Let $\varphi : \frakA \to \frakB$ be a surjective morphism of tree algebras
and suppose that $\frakA$~is regular. We have to show that $\frakB$~is also regular.
Fix a finite set $C \subseteq A$ of generators and set $D := \varphi[C]$.
Increasing~$C$ if necessary we may assume that $C = \varphi^{-1}[D]$.

First, note that $\frakB$~is finitely generated by~$D$.
Furthermore,
\begin{align*}
  t = \bbT\varphi(s)
  \qtextq{implies}
  \pi(t) = \pi(\bbT\varphi(s)) = \varphi(\pi(s))\,,
  \qquad\text{for } s \in \bbT C \text{ and } t \in \bbT D\,.
\end{align*}
Hence, for $b \in B_m$,
\begin{align*}
  \pi^{-1}(b) \cap \bbT D
  &= \set{ t \in \bbT D }{ \pi(t) = b } \\
  &= \bigset{ \bbT\varphi(s) }{ s \in \bbT C\,,\ \varphi(\pi(s)) = b } \\
  &= \bbT\varphi\bigl[
       \bigcup {\bigset{ \pi^{-1}(a) \cap \bbT C }{ a \in \varphi^{-1}(b) }}\bigr]
\end{align*}
Since $\varphi^{-1}(b) \subseteq A_m$ is a finite set,
the above union is finite and, therefore, regular.
As regular languages are closed under projections,
so is its image under~$\bbT\varphi$.
\end{proof}

As mentioned above, the definition of a regular tree algebra does not tell us
what these algebras look like. The next theorem sheds a bit more light on this
question. A less abstract characterisation will be given in
Section~\ref{Sect: deterministic}.
To state the theorem, we need the notion of a \emph{finitary sub-quotient}
of a tree algebra~$\frakA$. By definition this is an algebra which can be obtained
from a finitary subalgebra of~$\frakA$ by taking a quotient.
Recall that we say that a class~$\calC$ \emph{characterises
the regular languages} if a~language is regular if, and only if, it is
recognised by some algebra from~$\calC$.
\begin{thm}\label{Thm: abstract characterisation}
Let $\calC$~be an arbitrary class of finitary tree algebras that characterises
the regular languages and that is closed under finite products.
A finitary tree algebra~$\frakA$ is regular if, and only if, it is the rank-limit
of a sequence of finitary sub-quotients of algebras in~$\calC$.
\end{thm}
\begin{proof}
$(\Leftarrow)$
Let ${(\frakB^n)}_{n<\omega}$ be a sequence of algebras in~$\calC$
and let $\frakD^n$~be a sub-quotient of~$\frakB^n$ such that
${(\frakD^n)}_{n<\omega}$ converges to~$\frakA$.
As $\calC$ characterises the regular tree languages,
every algebra in~$\calC$ is regular.
Since the regular tree algebras are closed under finitely generated subalgebras
and quotients, it follows that each~$\frakD^n$ is regular.
Finally, so is the limit~$\frakA$
since the class of regular algebras is closed under rank-limits.

$(\Rightarrow)$
Suppose that $\frakA$~is regular.
Let $C \subseteq A$ be a finite set of generators and choose a number $k < \omega$
such that $C \subseteq A_{<k}$.
We construct a sequence ${(\frakB^n)}_{n<\omega}$ of algebras in~$\calC$
and sub-quotients $\frakD^n$~of~$\frakB^n$ such that
${(\frakD^n)}_{n<\omega}$ converges to~$\frakA$.

Let $n \leq \omega$.
For each $a \in A_{<n}$, we choose an algebra $\frakB_a \in \calC$
and a morphism $\varphi_a : \bbT C \to \frakB_a$ recognising
$\pi^{-1}(a) \cap \bbT C$.
Set
\begin{align*}
  \frakB^n := \prod_{a \in A_{<n}} \frakB_a
  \qtextq{and}
  \varphi := \langle\varphi_a\rangle_{a \in A_{<n}} : \bbT C \to \frakB^n\,.
\end{align*}
Let $\frakD' \subseteq \frakB^n$ be the subalgebra induced by the set
$D' := \rng \varphi$.
Note that $\frakD'$~is finitely generated by $\varphi[C]$.
We will show that
\begin{align*}
  \varphi(s) = \varphi(t) \quad\text{implies}\quad
  \pi(s) = \pi(t)\,, \quad\text{for } s,t \in \bbT_{<n} C\,.
\end{align*}
Then it follows by standard arguments that there exists a function
$\psi : D'_{<n} \to A_{<n}$ satisfying
$\psi \circ \varphi \restriction \bbT_{<n} C = \pi \restriction \bbT_{<n} C$.
As $\varphi$~and~$\pi$ are morphisms of $\bbT_{<n}$-algebras, so is~$\psi$.
And since $\pi \restriction \bbT_{<n} C$ is surjective, so is~$\psi$.
Consequently, $\psi : \frakD|_{<n} \to \frakA|_{<n}$ is a morphism.
Let $\frakD^n := \frakD'/{\ker \psi}$ where
$\ker \psi$ denotes the equivalence relation of `having the same image under~$\psi$'.
Then $\frakD^n$~is a sub-quotient of~$\frakB^n$
and $\frakD^n|_{<n} \cong \frakA|_{<n}$.
Consequently, ${(\frakD^n)}_{n<\omega}$ is a sequence
of finitary sub-quotients that converges (up to isomorphisms) to~$\frakA$.

It remains to prove the claim.
Let $s,t \in \bbT_{<n} C$ be trees with $\varphi(s) = \varphi(t)$.
By construction there exist sets $P_a \subseteq D'$, for $a \in A_{<n}$, such that
\begin{align*}
  \pi^{-1}(a) \cap \bbT C = \varphi^{-1}[P_a]\,.
\end{align*}
It follows that
\begin{align*}
  \pi(s) = a
  \quad\iff\quad \varphi(s) \in P_a
  \quad\iff\quad \varphi(t) \in P_a
  \quad\iff\quad \pi(t) = a\,.
\end{align*}
Hence, $\pi(s) = \pi(t)$, as desired.
\end{proof}

It follows by Theorems~\ref{Thm:regular algebra iff recognises regular languages} and~\ref{Thm:regular algebras characterise the regular languages}
that the class of regular tree algebras is the largest class that characterises
the regular languages.
From the preceding theorem we can now conclude that it is in fact the only
pseudo-variety with this property. This means that the notion of a
regular tree algebra is quite canonical, although we still would like to have a
more concrete definition.
\begin{cor}\label{Cor: uniqueness}
The class of regular tree algebras is the only pseudo-variety characterising
the class of regular tree languages.
\end{cor}
\begin{proof}
We have already shown that the class of regular tree algebras forms a pseudo-variety.
For uniqueness, let $\calC$~be any pseudo-variety characterising the regular tree
languages. Then every algebra in~$\calC$ is regular.
Conversely, let $\frakA$~be a regular tree algebra.
By Theorem~\ref{Thm: abstract characterisation}, there exist algebras
$\frakB_n \in \calC$ and finitary sub-quotients~$\frakD_n$ of~$\frakB_n$,
for $n < \omega$, such that $\frakA$~is the rank-limit of ${(\frakD_n)}_n$.
As $\calC$~is a pseudo-variety, it follows that every~$\frakD_n$ belongs to~$\calC$
and, therefore, also the limit~$\frakA$.
\end{proof}

Our definition of a pseudo-variety was complicated by the fact that the class of
finitary tree algebras is not closed under subalgebras and finite products.
Here we present two examples showing that a subalgebra or a finite product of regular tree algebras
need not be finitely generated.

(a) Let us start with subalgebras.
We use a result by Yanov and Muchnik~\cite{YanovMuchnik59} about so-called \emph{clones}.
A clone~$\frakC$ is a set of functions (of various arities) over some fixed set~$X$
that contains all projections and that is closed under composition, i.e.,
if $\frakC$~contains $f : X^n \to X$ and $g_0,\dots,g_{n-1} : X^m \to X$,
it also contains the $m$-ary function
\begin{align*}
  \bar x \mapsto f(g_0(\bar x),\dots,g_{n-1}(\bar x))\,.
\end{align*}
Note that this composition also makes sense if the functions $g_0,\dots,g_{n-1}$ have
different arities since we can make their arities equal by composing them by suitable
projections (which are in~$\frakC$ by assumption).
\begin{thm}[Yanov, Muchnik]
There are uncountably many clones on a three element set.
\end{thm}
As there are only countably many finitely generated clones, it follows in particular
that there exists some clone~$\frakC$ that is not finitely generated.
We will use it to construct the desired tree algebra.

Let $[3] = \{0,1,2\}$ be a three element set and let $A_n$~be the set of all
functions ${[3]}^n \to [3]$ together with a special error value~$\bot$.
We turn $A = {(A_n)}_n$ into a tree algebra by defining the following
multiplication $\pi : \bbT A \to A$.
For a finite tree $t \in \bbT A$ that does not contain the symbol~$\bot$,
we compute the product $\pi(t)$ by composing all the functions that label
the vertices of~$t$.
For all other trees, we set $\pi(t) := \bot$.
The resulting structure $\frakA = \langle A,\pi\rangle$
forms a tree algebra which is finitely generated.
(To see the latter, one can, e.g., represent every $3$-valued function in a similar way
as boolean functions can be written in disjunctive normal form.)
Furthermore, $\frakA$~is even regular since, when evaluating a tree~$t$
an automaton is able to first check that $t$~is finite and does not contain~$\bot$,
and then evaluate~$t$ bottom up by remembering where each (of the bounded number) of the
input arguments is mapped to.

To conclude the construction recall that we have seen above that there exists a clone on~$[3]$
that is not finitely generated.
Let $\frakC \subseteq \frakA$ be the subalgebra of~$\frakA$ consisting of the elements of that clone.
Then $\frakC$~is not finitely generated.

(b) Our counterexample for products looks as follows.
We start with a tree algebra~$\frakB$ where the elements of arity~$n$
are all finite sequences in ${\{ x_0,\dots,x_{n-1}\}}^*$ that contain every variable
at most once.
We define the product as follows.
Suppose we have sequences $\alpha \in B_m$ and $\beta_0,\dots,\beta_{m-1} \in B_n$
where the $\beta_i$~are disjoint. If $\alpha = \langle x_{i_0},\dots,x_{i_{k-1}}\rangle$,
we set
\begin{align*}
  \alpha(\beta_0,\dots,\beta_{m-1})
    := \beta_{i_0}\dots\beta_{i_{k-1}}\,,
\end{align*}
i.e., we substitute $\beta_i$~for~$x_i$ in~$\alpha$.
For a finite tree $t \in \bbT B_n$, we can now inductively define
\begin{align*}
  \pi(t) = \alpha(\pi(s_0),\dots,\pi(s_{m-1}))\,,
\end{align*}
where $\alpha := t(\emptyseq)$ is the label at the root and
$s_0,\dots,s_{m-1}$ are the attached subtrees.
(With the convention that $\pi(s_i) = \langle x_k\rangle$
in case that $s_i = x_k$ is a single variable.)

We can extend this definition to infinite trees as follows.
If $t$~does not contain variables, we set $\pi(t) = \emptyseq$.
Otherwise, we choose a finite prefix~$s$ of~$t$ that contains all the variables,
separately compute the products of $s$ and of the attached subtrees,
and then multiply the results as above.
Note that this definition ensures that $\pi(t)$~is the sequence
of all variables appearing in~$t$, but not necessarily in the order they appear in.

Again it is straightforward to check that $\frakB$~is a tree algebra.
Furthermore, note that we can write every sequence $\alpha \in B_m$
as the product of a tree~$t$ where all internal vertices are labelled by $\langle x_0\rangle$
or $\langle x_0,x_1\rangle$
by suitably choosing the ordering of the variables of~$t$.
Hence, $\frakB$~is finitely generated by three elements
$\emptyseq,\langle x_0\rangle,\langle x_0,x_1\rangle$.

Furthermore, $\frakB$~is regular since, given an element $b \in B_n$ and a finite set
of generators, an automaton can determine whether an input tree evaluates to~$b$
since all intermediate results are sequences of length at most~$n$.

We claim that the product $\frakB \times \frakB$ is not finitely generated.
For a contradiction suppose otherwise and fix a finite set $C$~of generators.
Choose a number~$m$ that is greater than the arity of all elements in~$C$.
We consider the element $\langle\alpha,\beta\rangle \in B_{2m} \times B_{2m}$
where
\begin{align*}
  \alpha &:= \langle x_0,\dots, x_{2m-1}\rangle \\
  \beta  &:= \langle x_m, x_0, x_{m+1}, x_1,\dots, x_{m+i},x_i,\dots, x_{2m-1}, x_{m-1}\rangle\,.
\end{align*}
By assumption, there is a tree~$t$ with product $\langle\alpha,\beta\rangle$.
Let $\langle\gamma,\delta\rangle$ be the label at the root of~$t$ and
let $s_0,\dots,s_{n-1}$ be the subtrees attached to it.
(For simplicity, we assume that $n > 1$. Otherwise our proof needs to be slightly modified.)
By choice of~$m$, there is some subtree~$s_i$ that contains at least two variables.
Let $\sigma,\tau : [n] \to [n]$ be the permutations such that
\begin{align*}
  \gamma = \langle x_{\sigma(0)},\dots,x_{\sigma(n-1)}\rangle
  \qtextq{and}
  \delta = \langle x_{\tau(0)},\dots,x_{\tau(n-1)}\rangle\,,
\end{align*}
and let $p : \frakB \times \frakB \to \frakB$ be the projection to the first component.
By looking at the first components, we see that
\begin{align*}
   \pi\bigl(\bbT p(s_{\sigma(0)})\bigr) \dots \pi\bigl(\bbT p(s_{\sigma(n-1)})\bigr)
  &= \gamma\bigl(\pi(\bbT p(s_0)),\dots,\pi(\bbT p(s_{n-1}))\bigr) \\
  &= \alpha = \langle x_0,\dots, x_{2m-1}\rangle\,.
\end{align*}
Consequently, there exist numbers $k < l$ such that
the term~$s_i$ contains the variables $x_k,x_{k+1},\dots,x_{l-1}$.
By choice of~$i$, we have $l \geq k+2$.

Looking at the second components, we see that $\beta$~must have some segment
of length $l - k \geq 2$ which contains the variables $x_k,x_{k+1},\dots,x_{l-1}$
(in any order).
But the only segments of~$\beta$ of this form are those of length~$1$
and the one of length~$2m$. A~contradiction.

\section{Syntactic algebras}%   %%%%%%%%%%%%%%%%%%%%%%%%%%%%%%%%%%%%%%%%%%%%%%%%%%%%%%%%%%%
\label{Sect: syntactic algebras}

Besides being a pseudo-variety we also need our class of recognisers to
have what is called \emph{syntactic algebras.}
These are algebras recognising a given language that are minimal in a certain sense.
Usually we can obtain such an algebra by taking a suitable quotient of the free algebra.
In this section we will show that for tree algebras the situation
is exactly the same. Let us start with some basic definitions.

A \emph{congruence} for a tree algebra~$\frakA$ is an equivalence
relation~$\approx$ on its universe~$A$ that is compatible with the product
in the sense that, if $s,t \in \bbT A$ are two trees of the same shape
such that $s(v) \approx t(v)$, for all~$v$, then $\pi(s) \approx \pi(t)$.
If $\approx$~is a congruence, we can define a tree algebra structure
on the quotient $A/{\approx}$ in the natural way.
We denote it by $\frakA/{\approx}$.

A \emph{tree with a hole,} or a \emph{context,} is a tree
$t \in \bbT(A \cup \Box)$ where the new symbol~$\Box$ is called the \emph{hole.}
It works as a kind of variable,
but with the difference that it can have an arbitrary (but fixed) arity and
that it can appear several times in~$t$.
Note that we allow~$\Box$ to have positive arity, which means that it can occur
in a non-leaf position in the tree.
Given such a context~$t$ and an element~$a$ of the right arity, we denote by
$t[a]$ the product $\pi(t')$ where $t'$~is the tree obtained from~$t$
by replacing all labels~$\Box$ by~$a$.
\begin{defi}
Let $\frakA$~be a tree algebra and $P \subseteq A_n$ a set of elements of
arity~$n$. The \emph{syntactic congruence} for~$P$ is defined by
\begin{align*}
  a \approx_P b \quad\defiff\quad t[a] \in P \Leftrightarrow t[b] \in P\,,
  \quad\text{for all contexts } t \in \bbT_n(A \cup \Box)\,.
\end{align*}
\end{defi}
The non-obvious part of this definition is the fact that the resulting equivalence
relation is indeed a congruence. In fact, the proof of the next result
crucially relies on the fact that the tree algebra in question is regular.
For arbitrary tree algebras the statement is simply false.
\begin{thm}\label{Thm:syntactic congruence is one}
The syntactic congruence on a regular tree algebra is a congruence.
\end{thm}

For the proof, we need to set up a bit of technical machinery.
Fix a finite ranked set~$\Sigma$,
let $\sim$~be an equivalence relation on~$\bbT\Sigma$,
and let $\calA$~and~$\calB$ be two non-deterministic parity automata.
We will define a game $\calG_{\sim}(\calA,\calB)$ where
the first player wins if, and only if, there exist two trees $S,T \in \bbT\bbT\Sigma$
of the same shape such that
\begin{itemize}
\item $S(v) \sim T(v)$, for all vertices $v$,
\item $\calA$~accepts $\Flat(S)$,
\item $\calB$~accepts $\Flat(T)$.
\end{itemize}
The game is a variant of the well-known Automaton--Pathfinder Game. % chktex 8
The only difference is that we simulate two automata at the same time
and that, instead of playing single letters, we play larger trees in each step.
The game has two players \emph{Automaton} and \emph{Pathfinder.}
Each round starts in a position of the form $\langle p,q\rangle$,
where $p$~is a state of~$\calA$ and $q$~one of~$\calB$.
We allow $p$~and/or~$q$ to be undefined, which we denote by the special symbol~$\bot$.
In the first round of the game, $p$~and~$q$ are the initial states of the respective
automata.
Given such a position $\langle p,q\rangle$,
\begin{itemize}
\item Automaton chooses two trees $s,t \in \bbT_m\Sigma$ with $s \sim t$ and $m < \omega$,
\item if $p \neq \bot$, he also chooses a profile~$\delta$ for some partial run of~$\calA$ on~$s$
  that starts in state~$p$, and
\item if $q \neq \bot$, he chooses a profile~$\varepsilon$ for some partial run of~$\calB$ on~$t$
  that starts in state~$q$.
\end{itemize}
Pathfinder responds by selecting a number $i < m$.
The \emph{outcome} of this round is the pair $\langle \delta|_i, \varepsilon|_i\rangle$ where
$|_i$~denotes the restriction to the $(i+1)$-th successor, that is,
\begin{align*}
  \langle r,k_0,p_0,\dots,k_{m-1},p_{m-1}\rangle\mathord{\bigm|}_i := \langle r,k_i,p_i\rangle\,.
\end{align*}
If there is no~$\delta$, we set $\delta|_i := \langle\bot,\bot,\bot\rangle$ and similarly
for~$\varepsilon$.
If this outcome is $\langle p,k,p'\rangle,\langle q,l,q'\rangle$,
the next round of the game will start in the position $\langle p',q'\rangle$.

If at some point in the game one of the players cannot make his choice,
that player loses the game.
Otherwise, the players produce an infinite sequence
$\langle \delta_0,\varepsilon_0\rangle,\langle \delta_1,\varepsilon_1\rangle,\dots$ of outcomes.
Let $k_i$~be the priority in~$\delta_i$ and $l_i$~the priority in~$\varepsilon_i$.
Player Automaton wins the game if each of the sequences $k_0,k_1,\dots$ and $l_0,l_1,\dots$
either satisfies the parity condition or if it contains the symbol~$\bot$.
Otherwise, Pathfinder wins.

Clearly, if there are two trees $S,T \in \bbT\bbT\Sigma$ of the same shape such that
\begin{itemize}
\item $S(v) \sim T(v)$, for all vertices $v$,
\item $\calA$~accepts $\Flat(S)$, and
\item $\calB$~accepts $\Flat(T)$,
\end{itemize}
then Automaton has the following winning strategy in $\calG_{\sim}(\calA,\calB)$.
He fixes two accepting runs $\varrho$~and~$\sigma$ on, respectively, $\Flat(S)$ and $\Flat(T)$.
During the game he descends through the trees $S$~and~$T$.
When the game reaches a vertex~$v$, Automaton chooses the trees $S(v)$~and~$T(v)$ and
the profiles of the subruns of $\varrho$~and~$\sigma$
that correspond to the trees $S(v)$~and~$T(v)$, respectively.

Conversely, if Automaton has a winning strategy in the game,
we can use it to construct
\begin{itemize}
\item two trees $S,T \in \bbT\bbT\Sigma$ such that $S(v) \sim T(v)$ for all~$v$ and
\item accepting runs of $\calA$~and~$\calB$ on, respectively, $\Flat(S)$ and $\Flat(T)$.
\end{itemize}

\begin{proof}[Proof of Theorem~\ref{Thm:syntactic congruence is one}]
Let $\frakA$~be a regular tree algebra and let $C \subseteq A$ be a finite set of generators.
For a contradiction, suppose that there exists a subset $P \subseteq A_m$
such that $\approx_P$~is not a congruence.
Then we can find two trees $s,t \in \bbT A$ (of the same shape) such that
\begin{align*}
  s(v) \approx_P t(v)\,, \quad\text{for all vertices } v\,,
  \quad\text{but}\quad
  \pi(s) \napprox_P \pi(t)\,.
\end{align*}
For every vertex~$v$, we can choose trees $S(v),T(v) \in \bbT C$ such that
$s(v) = \pi(S(v))$ and $t(v) = \pi(T(v))$.
This defines two trees $S,T \in \bbT\bbT C$ with $s = \bbT\pi(S)$ and
$t = \bbT\pi(T)$.

As the algebra~$\frakA$ is regular and every $\approx_P$-class~$[a]$ is finite
(by definition, $\approx_P$~only relates elements of the same arity),
we can construct automata~$\calA_a$, for $a \in A$, such that
\begin{align*}
  L(\calA_a) = \bigset{ t \in \bbT C }{ \pi(t) \approx_P a }
             = \bigcup_{b \,\approx_P\, a} (\pi^{-1}(b) \cap \bbT C)\,.
\end{align*}
Let $a := \pi(s)$ and $b := \pi(t)$.
We consider the game $\calG_{\approx_P}(\calA_a,\calA_b)$.
The trees $S$~and~$T$ show that Automaton has a winning strategy in this game.
As the winning condition of the game is regular, we can apply the
B\"uchi--Landweber Theorem, which tells us that Automaton even has a % chktex 8
finite-memory winning strategy.
Since the choice of $S(v)$~and~$T(v)$ by Automaton in the game only depends
on the current position $\langle p,q\rangle$ and on the contents of the memory,
there are only finitely many different values for $S(v)$ and $T(v)$.
This implies that there are only finitely many different labels
used by the trees $s$~and~$t$.
Consequently, we can get from~$s$ to~$t$ by a finite number of steps
in each of which we replace several occurrences of a single label of~$s$
by the corresponding label of~$t$.
Thus, there exists a sequence $u_0,\dots,u_n \in \bbT A$
such that $s = u_0$, $t = u_n$, and each $u_{i+1}$ is obtained from~$u_i$
by replacing (several occurrences of) a single label by an $\approx_P$-equivalent one, i.e.,
$u_i = r_i[a_i]$ and $u_{i+1} = r_i[b_i]$, for a suitable context
$r_i \in \bbT(A + \Box)$ and elements $a_i \approx_P b_i$ in~$A$.
By induction on~$i$, it now follows that $\pi(u_i) \approx_P \pi(s)$.
For $i = 0$, this is trivial\?; and for $i > 0$ it is sufficient to note that
$a_{i-1} \approx_P b_{i-1}$ implies
\begin{align*}
  p[\pi(u_{i-1})] = p[r_{i-1}[a_{i-1}]] \in L \ \Leftrightarrow\
  p[\pi(u_i)] = p[r_{i-1}[b_{i-1}]] \in L\,,
  \quad\text{for all contexts~$p$.}
\end{align*}
Consequently, we have $\pi(s) \approx_P \pi(t)$. A~contradiction.
\end{proof}

As a consequence we obtain the same statement for free algebras,
provided that the given subset is a regular language.
\begin{cor}\label{Cor:syntactic congruence for free algebra}
Let $L \subseteq \bbT_n\Sigma$ be a regular language.
Then the syntactic congruence for~$L$ is a congruence on $\bbT\Sigma$.
\end{cor}
\begin{proof}
Let $L \subseteq \bbT_n\Sigma$ be regular.
Then there exists a regular tree algebra~$\frakA$ and a morphism
$\varphi : \bbT\Sigma \to \frakA$ such that
$L = \varphi^{-1}[P]$ for some set $P \subseteq A_n$.
By Theorem~\ref{Thm:syntactic congruence is one}, the syntactic congruence $\approx_P$
of~$P$ is a congruence.
The claim now follows from two facts that are both straightforward to prove\?:
\begin{enumerate}[label={\normalfont(\arabic*)}]
\item $L = \varphi^{-1}[P]$ implies that
  \begin{align*}
    s \approx_L t \quad\text{iff}\quad \varphi(s) \approx_P \varphi(t)\,.
  \end{align*}
\item If $\sim$~is a congruence of~$\frakA$ then
  \begin{align*}
    a \sim_\varphi b \quad\defiff\quad \varphi(a) \sim \varphi(b)
  \end{align*}
  is a congruence of~$\bbT\Sigma$.
  \qedhere
\end{enumerate}
\end{proof}

\noindent
For a regular language $L \subseteq \bbT_n\Sigma$, we call the quotient
$\bbT\Sigma/{\approx_L}$ the \emph{syntactic algebra} of~$L$.
An immediate consequence of the way we have defined~$\approx_L$ is that
the syntactic algebra is minimal in the sense that
the projection $\bbT\Sigma \to \bbT\Sigma/{\approx_L}$ factorises through
every morphism $\bbT\Sigma \to \frakA$ that recognises~$L$.
\begin{thm}\label{Thm:syntactic algebras are universal}
The syntactic algebra of a regular tree language~$L$ is regular and it is
the smallest tree algebra recognising~$L$.
\end{thm}
\begin{proof}
Let $L \subseteq \bbT_n\Sigma$ be regular and let
$\varphi : \bbT\Sigma \to \frakA$ be a morphism recognising it.
Replacing $\frakA$ by the image of~$\varphi$ we may assume that $\varphi$~is surjective.
We start by constructing a morphism $\psi : \frakA \to \bbT\Sigma/{\approx_L}$
such that $\psi \circ \varphi = q$, where $q : \bbT\Sigma \to \bbT\Sigma/{\approx_L}$
is the quotient map.
\begin{center}
\includegraphics{Regular-submitted-6.mps}
%\begin{mpfig}
%  u := 2.3cm;
%
%  z0 = (0,0);
%  z1 = (u,0);
%  z2 = (u,-0.5u);
%
%  pickup pencircle scaled 0.6pt;
%
%  drawarrow anchor(z0,z1,7pt) -- anchor(z1,z0,7pt);
%  drawarrow anchor(z0,z2,7pt) -- anchor(z2,z0,15pt);
%  drawarrow anchor(z1,z2,7pt) -- anchor(z2,z1,7pt);
%
%  label (btex $\bbT\Sigma$ etex, z0);
%  label (btex $\frakA$ etex, z1);
%  label (btex $\bbT\Sigma/{\approx_L}$ etex, z2);
%
%  label top  (btex $\varphi$ etex, 1/2[z0,z1]);
%  label llft (btex $q$ etex,       1/2[z0,z2]);
%  label rt   (btex $\psi$ etex,    1/2[z1,z2]);
%\end{mpfig}
\end{center}
To do so it is sufficient to prove that
\begin{align*}
  \varphi(s) = \varphi(t)
  \qtextq{implies}
  s \approx_L t\,.
\end{align*}
Then we can define $\psi(a) := q(t)$, for some $t \in \varphi^{-1}(a)$.
By the above implication, $\psi$~is well-defined.
Furthermore, it is straightforward to check that this function is in fact a morphism
of tree algebras.

Hence, it remains to prove the claim.
Suppose that $\varphi(s) = \varphi(t)$. To show that $s \approx_L t$
consider a context~$r$ with $r[s] \in L$.
Then $\varphi(r[s]) \in P := \varphi[L]$.
Let $r' := \bbT\varphi(r)$. Then
\begin{align*}
  \varphi(r[t]) = r'[\varphi(t)] = r'[\varphi(s)] = \varphi(r[s]) \in P\,,
\end{align*}
which implies that $r[t] \in L$, as desired.

To conclude the proof, we have to show that the syntactic algebra
$\bbT\Sigma/{\approx_L}$ is regular.
Fix a morphism $\varphi : \bbT\Sigma \to \frakA$ into a regular algebra
recognising~$L$. We have just shown that there exists a morphism
$\psi : \frakA \to \bbT\Sigma/{\approx_L}$ with $\psi \circ \varphi = q$.
As the quotient map~$q$ is surjective, so is~$\psi$.
Hence, $\bbT\Sigma/{\approx_L}$ is a quotient of~$\frakA$ and,
therefore, regular by Theorem~\ref{Thm:regular algebras are pseudo-variety}.
\end{proof}

We have just proved the existence of syntactic algebras in our framework.
If we want to use our theory to develop decidable characterisations
of logical fragments, we further require an algorithm to actually compute
these algebras.
Before presenting one we need to explain how to represent a regular tree algebra
to an algorithm. The problem is that, while finitary, a regular tree algebra
still has infinitely many elements. So we cannot simply write down its
multiplication table.
What we do instead is to use an algorithm that, given an arity $n < \omega$,
produces a (finite) list of automata, one for each language of the form
$\pi^{-1}(a) \cap \bbT C$ for $a \in A_n$.
Using this representation, we can then algorithmically construct and process regular
tree algebras.
\begin{thm}\label{Thm: syntactic algebras are computable}
Given a regular language $L \subseteq \bbT_n\Sigma$,
we can compute the syntactic algebra $\bbT\Sigma/{\approx_L}$.
\end{thm}
\begin{proof}
Let $L \subseteq \bbT_n\Sigma$ be regular and $\calA$~an automaton for~$L$.
Using the construction from the proof of
Theorem~\ref{Thm:regular algebras characterise the regular languages},
we can compute a regular tree algebra~$\frakA$,
a morphism $\varphi : \bbT\Sigma \to \frakA$, and a set
$P \subseteq A_n$ such that $L = \varphi^{-1}[P]$.
By Theorem~\ref{Thm:syntactic algebras are universal}, it follows that
\begin{align*}
  \bbT\Sigma/{\approx_L} \cong \frakA/{\approx_P}\,.
\end{align*}
Hence, it is sufficient to prove that the relation~$\approx_P$ is decidable.
Note that
\begin{align*}
  a \napprox_L b
  \quad\iff\quad \text{there exists some context } t \text{ with }
                 (t[a] \in L \Leftrightarrow t[b] \notin L)\,.
\end{align*}
We will prove the decidability of the latter condition.

Let $C \subseteq A$ be a finite set of generators of~$\frakA$.
W.l.o.g.\ we may assume that all labels of the term~$t$
we are looking for (except for the hole~$\Box$) are in~$C$.
As $\pi^{-1}(a) \cap \bbT C$ is regular,
there exists a regular tree $u \in \bbT C$ with $\pi(u) = a$.
Similarly, we can find a regular tree $v \in \bbT C$ with $\pi(v) = b$.
Let $m$~be the arity of $a$~and~$b$ and
fix finite graphs $G$~and~$H$ whose unravellings are, respectively, $u$~and~$v$.
Given $G$~we can compute the set~$U$ of all tuples
$\langle p,q_0,\dots,q_{m-1}\rangle$ such that there exists a partial run~$\varrho$ of~$\calA$
on the tree~$u$ such that
\begin{itemize}
\item $\varrho$ starts in state~$p$,
\item the leaf with the variable~$x_i$ has state~$q_i$, and
\item every infinite branch satisfies the parity condition.
\end{itemize}
Similarly, we can compute an analogous set~$V$ for the tree~$v$.
Given these two sets we can then construct an automaton~$\calB$
that reads a context~$s$ and checks whether
the original automaton~$\calA$ accepts the tree $s[u]$, but does not
accept $s[v]$, or vice versa.
It follows that
\begin{align*}
  a \napprox_P b \quad\iff\quad L(\calB) \neq \emptyset\,,
\end{align*}
a condition that is decidable.
\end{proof}

As an example of how to use syntactic algebras let us derive a characterisation of
the class of \emph{commutative} tree languages.
For space reasons, we defer more substantial applications to a forthcoming
article~\cite{BlumensathYY}.
We say that a tree~$s$ is a \emph{permutation} of the tree~$t$ if $s$~is obtained
from~$t$ by rearranging the successors of every vertex.
Formally, we call a function $\sigma : \dom(s) \to \dom(t)$ a \emph{permutation}
if it is bijective and it preserves the successor and sibling relations.
Then $s$~is a \emph{permutation} of~$t$ if there exists some permutation
$\dom(s) \to \dom(t)$. A language $L \subseteq \bbT\Sigma$ is \emph{commutative}
if it is closed under permutations.
Note that this is not the same as saying that $L$~is closed under rearranging
the successors of a single vertex (or finitely many of them).
\begin{thm}
A regular tree language $L \subseteq \bbT_n\Sigma$ is commutative if, and only if,
its syntactic algebra~$\frakA$ satisfies the equations
\begin{align*}
  a(x_0,\dots,x_{m-1}) = a(x_{\sigma(0)},\dots,x_{\sigma(m-1)})\,,
\end{align*}
for all $a \in A_m$, $m < \omega$, and all permutations $\sigma : [m] \to [m]$.
\end{thm}
\begin{proof}
$(\Leftarrow)$
Note that the quotient morphism $\varphi : \bbT\Sigma \to \frakA$ recognises~$L$.
If $s$~is a permutation of~$t$, we have $\varphi(s) = \varphi(t)$.
Hence, $s \in L \Leftrightarrow t \in L$ and $L$~is commutative.

$(\Rightarrow)$
Fix an element $a \in A_m$ and a permutation $\sigma : [m] \to [m]$.
We have to show that
\begin{align*}
  a(x_0,\dots,x_{m-1}) \approx_L a(x_{\sigma(0)},\dots,x_{\sigma(m-1)})\,.
\end{align*}
Hence, let $r$~be a context. Note that the two trees obtained from~$r$
by replacing the hole~$\Box$ by, respectively, $a(x_0,\dots,x_{m-1})$ and
$a(x_{\sigma(0)},\dots,x_{\sigma(m-1)})$ are permutations of each other.
As $L$~is commutative we therefore have
\[
  r[a(x_0,\dots,x_{m-1})] \in L \Leftrightarrow
  r[a(x_{\sigma(0)},\dots,x_{\sigma(m-1)})] \in L\,.
  \qedhere
\]
\end{proof}

Note that it follows in particular that commutativity is decidable.
Given a regular language~$L$, we can compute its syntactic algebra and check
whether it satisfies the above equations. (We only need to check
them for elements~$a$ in a finite set of generators.)

\section{Deterministic tree algebras}%   %%%%%%%%%%%%%%%%%%%%%%%%%%%%%%%%%%%%%%%%%%%%%%%%%%
\label{Sect: deterministic}

In Theorem~\ref{Thm: abstract characterisation} we have provided a characterisation of
regular tree algebras in terms of an unspecified second class~$\calC$ of algebras that
characterises the regular languages. We can obtain a more informative result by
making a concrete choice for~$\calC$. In this section we will consider one such class.
In order to make this second class as simple as possible, we allow the relationship
between the two classes to be more complicated that what we had in
Theorem~\ref{Thm: abstract characterisation}.

In addition, the characterisation we obtain in Theorem~\ref{Thm: second characterisation}
below can also serve as an alternative definition of the notion of a regular tree algebra.
It has the advantage that it is purely algebraic and does not refer to automata,
avoiding the apparent circularity of the original definition.
Its main disadvantage is that it is more complicated and abstract,
which is why we did not adopt it as the official definition.

For the definition, we need to work with \emph{ordered} algebras.
An ordered tree algebra $\frakA = \langle A,\pi,{\leq}\rangle$ consists of a
tree algebra $\langle A,\pi\rangle$ that is expanded by a partial order~$\leq$
on~$A$ such that the product~$\pi$ is monotone.
(We order $\bbT A$ componentwise\?: $s \leq t$ if the trees $s$~and~$t$ have the
same shape and each label of~$s$ is less than or equal to the corresponding label
of~$t$.)
Such an ordering is \emph{complete} if it has arbitrary joins and meets
(i.e., suprema and infima).
Morphisms of ordered tree algebras are assumed to preserve the ordering and
morphisms of completely ordered algebras are assumed to also preserve joins and meets.
The class of algebras we are considering in this section is the following one.
\begin{defi}
Let $\frakA$~be a completely ordered tree algebra.

(a) $\frakA$~is \emph{meet-continuous} if products commute with meets, that is,
given a tree $T \in \bbT\PSet(A)$ labelled by subsets of~$A$, we have
\begin{align*}
  \pi(\bbT\inf(T)) = \inf {\set{ \pi(t) }{ t(v) \in T(v) \text{ for all vertices } v }}\,.
\end{align*}

(b) An element $a \in A_n$ is \emph{rectangular} if it can be written as a meet
of elements of arity~$0$ and elements of the form $b(x_i)$, for $b \in A_1$ and $i < n$.

(c) $\frakA$~is \emph{deterministic} if
it is meet-continuous and all elements are rectangular.
\end{defi}

\noindent
The motivating example for a deterministic algebra is one arising from an automaton
in the following way.
\begin{defi}
Let $\calA$~be a tree automaton and let $\frakA$~be the
tree algebra constructed at the end of Section~\ref{Sect: regular algebras}.
The \emph{transition algebra}~$\frakT(\calA)$ of~$\calA$ is the subalgebra of~$\frakA$
whose elements are
conjunctions of semigroup elements plus the empty disjunction~$\bot$,
i.e., we omit all disjunctions with more than one term.
We consider $\frakT(\calA)$ an ordered algebra where the ordering
is the one induced by the conjunctions and disjunctions.
\end{defi}
\begin{lem}\label{Lem: transition algebras are deterministic}
The transition algebra $\frakT(\calA)$ is deterministic.
\end{lem}
\begin{proof}
By definition, every element is a meet (conjunction) of elements of arity~$0$
or elements of the form $a(x_i)$ where $a$~has arity~$1$.
Thus, all elements are rectangular.
For meet-continuity it is sufficient to note that,
in every tree algebra constructed from an $\omega$-semigroup as in the proof of
Theorem~\ref{Thm:regular algebras characterise the regular languages},
the subalgebra consisting of the one-element disjunctions is meet-continuous
(see Proposition~4.12\,(a) of~\cite{BlumensathXX}).
\end{proof}

Deterministic algebras are a very special case of regular tree algebras.
One can show that their expressive power corresponds to a certain form of
deterministic tree automata.
For our purposes, it is sufficient to know that they are regular.
\begin{prop}\label{Prop: deterministic algebras are regular}
Every finitary subalgebra of a deterministic tree algebra is regular.
\end{prop}
\begin{proof}
Let $t$~be a tree we want to multiply.
As every label of~$t$ is rectangular, we can use meet-continuity
to transform the product of~$t$ into a meet of products where every
label has arity at most one. Such products correspond to $\omega$-semigroup
products along a single branch of~$t$
(see Lemma~4.23\,(b) of~\cite{BlumensathXX}).
This is something an automaton can evaluate.
Consequently, in order to check whether $t$~evaluates to a given element~$a$
an automaton can compute all the products along the branches of~$t$,
take their infimum, and compare it to~$a$.
\end{proof}

Let us use deterministic algebras to give a second characterisation
of the regular algebras.
We start with an observation that simplifies proofs of regularity\?:
we only need to check elements of arity at most one.
\begin{prop}\label{Prop: reduction for arities}
A finitary tree algebra~$\frakA$ is regular if, and only if, it has a finite
set $C \subseteq A$ of generators such that
\begin{align*}
  \pi^{-1}(a) \cap \bbT C \text{ is regular\,,}
  \quad\text{for every } a \in A \text{ of arity at most } 1\,.
\end{align*}
\end{prop}
Before giving the proof, we need to collect a few results about factorisations.
A \emph{factorisation} of a tree $t \in \bbT A$
is a tree $T \in \bbT\bbT A$ such that $\Flat(T) = t$.
We denote by $\calF(t)$ the set of all factorisations~$T$ of~$t$ such that
the trees $T(v)$ are singletons for all vertices~$v$ of~$T$ with more than one
successor.
The \emph{height} of a factorisation~$T$ is the height of the tree~$T$.

We call a tree $t \in \bbT A$ \emph{reduced} if it has no non-trivial factor of arity at most
one, that is, for every factorisation~$T$ of~$t$ and every vertex $v \in \dom(T)$
of arity at most one, we have $T(v) = \sing(a)$, for some $a \in A$.
The important fact about reduced trees is that they are small.
\begin{lem}\label{Lem: reduced trees are small}
Let $\frakA$~be a tree algebra and $m < \omega$.
Every reduced tree $t \in \bbT_m A$ has height at most~$2m$.
\end{lem}
\begin{proof}
We prove the claim by induction on~$m$.
For $m = 0$, note that every reduced tree of arity~$m$ is of the form $\sing(a)$,
for some $a \in A$. Hence, the height is~$0$.
For the inductive step, suppose that $m > 0$ and consider a reduced tree $t \in \bbT_m A$.
We distinguish two cases.

First, suppose that the root has an arity greater than~$1$.
As~$t$ is reduced, every subtree attached to the root must have fewer variables
than~$t$. By inductive hypothesis, their height is at most $2(m-1)$.
Hence, the height of $t$~is at most $2(m-1) + 1$.

It remains to consider the case where the root has arity~$1$.
As $t$~is reduced, the successor must then have arity greater than~$1$.
Hence, the attached subtree satisfies the above case, which means that its
height is bounded by $2(m-1) + 1$. Consequently, the height of~$t$ is at most
$2(m-1) + 2 = 2m$.
\end{proof}

Next we will show that the set $\calF(t)$ of factorisations of~$t$
contains reduced trees.
For the proof we will employ the following ordering on~$\calF(t)$.
For $S,T \in \calF(t)$, we set
\begin{align*}
  S \sqsubseteq T
  \quad\defiff\quad
  &\text{there is some } U \in \bbT\bbT\bbT A \text{ such that }
   S = \Flat(U) \text{ and} \\
  &\text{every } U(v) \text{ is a factorisation of } T(v)\,,
   \text{ for } v \in \dom(U)\,.
\end{align*}
\begin{lem}\label{Lem: F(t) inductively ordered}
The set $\calF(t)$ is inductively ordered by~$\sqsubseteq$, i.e.,
every chain as an upper bound.
\end{lem}
\begin{proof}
Let ${(T_i)}_{i\in I}$ be an increasing sequence in~$\calF(t)$.
We have to find an upper bound.
Note that every factorisation~$T$ of~$t$ induces an equivalence relation~$\approx_T$
on $\dom(t)$ by
\begin{align*}
  u \approx_T v \quad\defiff\quad
  u \text{ and } v \text{ are vertices belonging to the same factor } T(w)\,.
\end{align*}
Hence, the sequence $T_0 \sqsubseteq T_1 \sqsubseteq T_2 \sqsubseteq \dots$
induces a corresponding sequence
${\approx_{T_0}} \subseteq {\approx_{T_1}} \subseteq {\approx_{T_2}} \subseteq\dots$
of equivalence relations. The limit
\begin{align*}
  {\approx} := \bigcup_{i \in I} {\approx_i}
\end{align*}
is an equivalence relation on $\dom(t)$ that corresponds to some factorisation~$T$
of~$t$.
We will show that $T \in \calF(t)$.
Then $T$~is the desired upper bound for ${(T_i)}_{i \in I}$.

To prove the claim, note that every $\approx$-class~$E$ is the union of an increasing
sequence ${(E_i)}_{i \in I}$ of ${\approx_{T_i}}$-classes.
Since each~$T_i$ belongs to $\calF(t)$, every $E_i$ is of one of the following two
types.
\begin{enumerate}[label={\normalfont(\Roman*)}]
\item The class is a singleton.
\item The class corresponds to a factor of arity at most one.
\end{enumerate}
If there are arbitrarily large~$i$ such that $E_i$~is of type~(I), the sequence
is constant and the limit~$E$ is also of type~(I).
Otherwise, the limit~$E$ is a union of classes of type~(II) and, hence, is also
of type~(II).
As this holds for all classes of~$\approx$, it follows that $T \in \calF(t)$.
\end{proof}

\begin{lem}\label{Lem: existence of reduced factorisation}
Let $\frakA$~be a tree algebra and $C \subseteq A$ a set with
$A_0 \cup A_1 \subseteq C$.
Every $t \in \bbT_m C$ has a factorisation $T \in \calF(t)$ such that
\begin{enumerate}[label={\normalfont(\arabic*)}]
\item $T$~is reduced,
\item the height of~$T$ is at most $2m$, and
\item $\bbT\pi(T) \in \bbT C$.
\end{enumerate}
\end{lem}
\begin{proof}
By Lemma~\ref{Lem: F(t) inductively ordered}, we can use Zorn's Lemma to find a
maximal element $T \in \calF(t)$.
We claim that~$T$ is the desired factorisation.

(1) For a contradiction, suppose otherwise.
Then there exists a factorisation~$U$ of~$T$ and a vertex $u \in \dom(U)$ of arity at
most one such that $U(u)$ is not a singleton.
Let $T'$~be the tree obtained from~$T$ by replacing the factor~$U(u)$ by its product.
Then, $T \sqsubset T'$ and $T$~is not maximal.

(2) follows from (1) by Lemma~\ref{Lem: reduced trees are small}.

(3) Note that every factor $T(v)$ is either a singleton or of arity at most one.
Since $A_0 \cup A_1 \subseteq C$, it follows that $\pi(T(v)) \in C$.
Hence, $\bbT\pi(T) \in \bbT C$.
\end{proof}

\begin{proof}[Proof of Proposition~\ref{Prop: reduction for arities}]
For the nontrivial direction,
suppose that $\frakA$~is an algebra as in the proposition and
let $C \subseteq A$ be the corresponding set of generators.
To prove that $\frakA$~is regular, we fix an element $a \in A_m$.
We have to show that $\pi^{-1}(a) \cap \bbT C$ is regular.
Set $C' := C \cup A_0 \cup A_1$ and let $t \in \bbT C$.
By Lemma~\ref{Lem: existence of reduced factorisation},
$t$~has a factorisation $T \in \calF(t)$ such that $T$~is reduced, its height
is at most~$2m$, and $\bbT\pi(T) \in \bbT C'$.
It follows that $\bbT\pi(T) \in H(a)$ where
\begin{align*}
  H(a) := \set{ s \in \bbT C' }
              { s \text{ has height at most } 2m \text{ and } \pi(s) = a }\,.
\end{align*}
Consequently, we have
\begin{align*}
  \pi(t) = a
  \quad\iff\quad
  \pi(\Flat(T)) = a
  \quad\iff\quad
  \pi(\bbT\pi(T)) = a
  \quad\iff\quad
  \bbT\pi(T) \in H(a)\,.
\end{align*}
For every finite tree~$s$, we will construct an $\MSO$-formula~$\vartheta_s$ such that
\begin{align*}
 t \models \vartheta_s \quad\iff\quad
 t \text{ has a factorisation $T \in \calF(t)$ such that } \bbT\pi(T) = s\,.
\end{align*}
Then it follows that
\begin{align*}
  \pi(t) = a
  \quad\iff\quad
  \bbT\pi(T) \in H(a)
  \quad\iff\quad
  t \models \Lor_{s \in H(a)} \vartheta_s\,,
\end{align*}
as desired. Hence, it remains to construct the formulae~$\vartheta_s$.

First, note that we can encode a factorisation~$T$ of~$t$ by a set~$Z$
that contains the root of each factor~$T(v)$. Using this encoding, we can set
\begin{align*}
  \vartheta_s := \exists Z\Bigl[&\text{`$Z$ encodes a factorisation $T$ in $\calF(t)$'} \\
    \land &\Land_{v \in \dom(s)} \text{`the factor $T(v)$ evaluates to $s(v)$'}\Bigr]\,.
\end{align*}
The first part of this formula is clearly expressible in $\MSO$.
For the second part, note that $s$~is finite and each factor~$T(v)$ is either
a singleton or a term of arity at most one.
In the first case it is trivial to compute the product.
In the second case, we can use the formulae defining the sets
$\pi^{-1}(a) \cap \bbT C$, for $a \in A_0 \cup A_1$.
\end{proof}

The price we pay for using deterministic algebras in our characterisation theorem below
is that we need a slightly more general notion of recognition.
A~\emph{span} $\langle\varphi,\psi\rangle : \frakA \to \frakB$ from a tree algebra~$\frakA$
to another tree algebra~$\frakB$ consists of two morphisms
$\varphi : \frakC \to \frakA$ and $\psi : \frakC \to \frakB$
where $\frakC$~is a third tree algebra.
A subset $L \subseteq A_n$ is \emph{recognised} by a span $\langle\varphi,\psi\rangle$
if there exists a set $P \subseteq B_n$ such that
\begin{align*}
  L = \varphi[\psi^{-1}[P]]\,.
\end{align*}
Below we will use a span $\langle p,q\rangle : \frakA \to \frakT(\calA)$
where the middle algebra is a subalgebra of the product $\frakA \times \frakT(\calA)$
and the morphisms $p$~and~$q$ are the corresponding projections.
\begin{defi}\label{Def: hat A}
Let $\frakA$~be a tree algebra and $\calA$~an automaton.
We denote by $\widehat\frakA$~the subalgebra of the product
$\frakA \times \frakT(\calA)$ with domains
\begin{align*}
  \widehat A_n :=
    \bigset{ \langle \pi(t),\delta\rangle }
           { t \in \bbT_{n}A \text{ and }
             \delta \text{ the profile of some partial run of } \calA \text{ on } t }\,.
\end{align*}
Let $p : \widehat\frakA \to \frakA$ and $q : \widehat\frakA \to \frakT(\calA)$
be the corresponding projections.
\end{defi}
(Note that $\widehat\frakA$~is well defined as its domains are closed
under products.)
We start with a technical result showing that
the projection $\widehat\frakA \to \frakA$~is surjective.
At least this is the case if the algebra~$\frakA$ is regular and $\calA$~the corresponding
automaton, i.e., a~tree automaton such that, for every element $a \in A_0 \cup A_1$,
we can choose a starting state for~$\calA$ from which it recognises the set
$\pi^{-1}(a) \cap \bbT C$.
\begin{lem}\label{Lem: basic properties of hat A}
Let $\frakA$~be a regular tree algebra and $\calA$~an automaton for~$\frakA$.
The projection $p : \widehat\frakA \to \frakA$
is surjective and every fibre $p^{-1}(a)$~is finite.
\end{lem}
\begin{proof}
Consider an element $a \in A_m$.
Let $\varrho$~be the run of~$\calA$ on the tree $\sing(a)$ and
$\delta$~the (profile corresponding to the) transition at the root of~$\varrho$.
Then $\langle a,\delta\rangle \in \widehat A$ and
$p(\langle a,\delta\rangle) = a$.
Hence, $a \in \rng p$.
For the second statement, note that every domain $\widehat A_m \subseteq A_m \times T_m(\calA)$
is finite. Hence, so is $p^{-1}(a) \subseteq \widehat A_m$, for $a \in A_m$.
\end{proof}

Combining the notions and results of this section, we obtain the following characterisation
of when a tree algebra is regular.
\begin{thm}\label{Thm: second characterisation}
Let $\frakA$~be a finitary tree algebra and $C \subseteq A$ a finite set of generators.
$\frakA$~is regular if, and only if,
there exists a deterministic algebra~$\frakD$ and
a subalgebra $\widehat\frakA \subseteq \frakA \times \frakD$
such that

\noindent
\begin{minipage}[t]{0.73\textwidth}
\begin{itemize}
\item the first projection $p : \widehat\frakA \to \frakA$ is surjective,
\item every fibre $p^{-1}(a)$~is finite, and
\item the span $\langle\bbT p,\pi \circ \bbT q\rangle : \bbT\frakA \to \frakD$
  recognises every preimage
  \begin{align*}
    \pi^{-1}(a) \cap \bbT\,, \quad\text{for } a \in A_0 \cup A_1\,.
  \end{align*}
\end{itemize}
\end{minipage}%
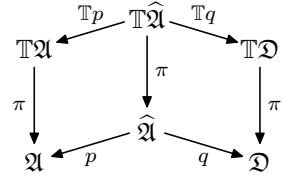
\begin{minipage}[t]{0.27\textwidth}
\vspace*{0pt}
\vspace*{-2ex}
\includegraphics{Regular-submitted-7.mps}
%\begin{mpfig}
%  u := 1.5cm;
%
%  z0 = (0,0);
%  z1 = (0,u);
%  z2 = (-u,-0.3u);
%  z3 = (-u,0.7u);
%  z4 = ( u,-0.3u);
%  z5 = ( u,0.7u);
%
%  pickup pencircle scaled 0.6pt;
%
%  drawarrow anchor(z1,z0,8pt) -- anchor(z0,z1,8pt);
%  drawarrow anchor(z3,z2,7pt) -- anchor(z2,z3,7pt);
%  drawarrow anchor(z5,z4,7pt) -- anchor(z4,z5,7pt);
%  drawarrow anchor(z0,z2,7pt) -- anchor(z2,z0,7pt);
%  drawarrow anchor(z0,z4,7pt) -- anchor(z4,z0,7pt);
%  drawarrow anchor(z1,z3,9pt) -- anchor(z3,z1,9pt);
%  drawarrow anchor(z1,z5,9pt) -- anchor(z5,z1,9pt);
%
%  label (btex \normalsize $\widehat\frakA$     etex, z0);
%  label (btex \normalsize $\bbT\widehat\frakA$ etex, z1);
%  label (btex \normalsize $\frakA$             etex, z2);
%  label (btex \normalsize $\bbT\frakA$         etex, z3);
%  label (btex \normalsize $\frakD$             etex, z4);
%  label (btex \normalsize $\bbT\frakD$         etex, z5);
%
%  label rt  (btex $\pi$    etex, 1/2[z1,z0]);
%  label lft (btex $\pi$    etex, 1/2[z3,z2]);
%  label rt  (btex $\pi$    etex, 1/2[z5,z4]);
%  label bot (btex $p$      etex, 1/2[z0,z2]);
%  label bot (btex $q$      etex, 1/2[z0,z4]);
%  label top (btex $\bbT p$ etex, 1/2[z1,z3]);
%  label top (btex $\bbT q$ etex, 1/2[z1,z5]);
%\end{mpfig}
\end{minipage}
\end{thm}
\begin{proof}
$(\Rightarrow)$
Fix an automaton~$\calA$ for~$\frakA$, set $\frakD := \frakT(\calA)$,
and let $\widehat\frakA$~be the algebra from Definition~\ref{Def: hat A}.
We have seen above that $\frakD$~is deterministic, the projection
$p : \widehat\frakA \to \frakA$ is surjective, and all fibres $p^{-1}(a)$ are finite.
To conclude the proof, consider an element $a \in A_0 \cup A_1$.
Let $q_a$~be the starting state that $\calA$~uses to recognise the preimgae
$\pi^{-1}(a) \cap \bbT C$ and let $P \subseteq D$ be the set of all profiles
$\Land_{i<m} \langle q,k_i,p_i\rangle$ such that
$q = q_a$ and, from the state~$p_i$, $\calA$~accepts the singleton tree with label~$x_i$,
for $i < m$.
For $t \in \bbT C$, it follows that
\begin{align*}
  \pi(t) = a
  &\quad\iff\quad
   \text{there exists an accepting run on } t \text{ starting in the state } q_a \\
  &\quad\iff\quad
   \text{there exists } s \in {(\bbT p)}^{-1}(t) \text{ such that } \bbT q(s)
   \text{ is such a run} \\
  &\quad\iff\quad
   \text{there exists } s \in {(\bbT p)}^{-1}(t) \text{ such that }
   \pi(\bbT q(s)) \in P\,.
\end{align*}

$(\Leftarrow)$
By Proposition~\ref{Prop: reduction for arities}, it is sufficient to show that
the preimages $\pi^{-1}(a) \cap \bbT C$ are regular for elements~$a$ of arity at most~$1$.
Hence, let $a \in A_0 \cup A_1$ and set $C' := q[p^{-1}[C]]$.
Note that $C'$~is a finite set since, by assumption, all fibres of~$p$ are finite.
Furthermore, we know that there exists a (finite) set $P \subseteq D$ such that
\begin{align*}
  \pi^{-1}(a) \cap \bbT C
  &= (\bbT p)[{(\pi \circ \bbT q)}^{-1}[P]\bigr] \\
  &= \bigcup_{d \in P} {(\bbT p)\bigl[{(\bbT q)}^{-1}[\pi^{-1}(d)]\bigr]} \\
  &= \bigcup_{d \in P} {(\bbT p)\bigl[{(\bbT q)}^{-1}[\pi^{-1}(d) \cap C']\bigr]}\,,
\end{align*}
where the last equality holds as every tree in $\bbT C$ is mapped by $\bbT q \circ {(\bbT p)}^{-1}$
to a tree in $\bbT C'$.
As finitary subalgebras of deterministic algebras are regular, each preimage $\pi^{-1}(d) \cap C'$
forms a regular language. Furthermore, regular tree languages are closed
under projections and inverse projections. Hence,
each term in the above union is regular and, therefore, so is the union itself.
\end{proof}

\section{Conclusion}   %%%%%%%%%%%%%%%%%%%%%%%%%%%%%%%%%%%%%%%%%%%%%%%%%%%%%%%%%%%%%%%%%%%%

In this article we have developed a framework for recognisability of tree languages.
We have isolated a class of algebras that recognise exactly the regular tree languages
and we have shown that this class meets our main requirements\?: it forms a pseudo-variety
and it has syntactic algebras. Furthermore, we have proved that it is the only class
with these properties. Finally, we have included a simple example of how to use our
framework to obtain characterisation results.
More substantial applications are deferred to a forthcoming article~\cite{BlumensathYY}.

The basic concept our framework is built around is the notion of a regular tree algebra.
We have given two different definitions of these algebras\?:
the first one is simple and easy to use, but it requires automata theory\?;
the second one is more complicated and abstract, but it has the
advantage that it is purely algebraic and does not require automata.
It is currently open whether one can also define regularity of a tree algebra
in terms of a set of equations. The work of Milius and Urbat~\cite{MiliusUrbat19}
suggests that this might be possible, but no explicit description of the equations involved
is known at this point. It is also unclear how large and complicated such a set of equations
would be. Our current conjecture is that, for every finite set~$X$ of variables
and every tree $t \in \bbT X$, we need to have an equation of the form $t = t'$,
where $t' \in \bbT X$ is some \emph{regular} tree depending on~$t$.
But it is not obvious why this should be equivalent to the tree algebra being regular.
In fact, both directions of this equivalence seem to require non-trivial arguments.

\appendix
\section{Closure properties of regular tree languages}   %%%%%%%%%%%%%%%%%%%%%%%%%%%%%%%%%%

The closure properties of the class of all regular languages of infinite trees is well-understood.
In particular, the class is closed under boolean operations and projections
(see, e.g.,~\cite{Thomas90,Thomas97,Niesser02}).
Another well-known closure property is that under inverse morphisms.
As~I~have not been able to find a published proof of this fact, I~include one here.
\begin{lem}\label{Lem: closure under inverse morphisms}
Let $\varphi : \bbT\Sigma \to \bbT\Gamma$ be a morphism of tree algebras.
If $L \subseteq \bbT_m\Gamma$ is regular, so is $\varphi^{-1}[L] \subseteq \bbT_m\Sigma$.
\end{lem}
\begin{proof}
Set $\widehat\Sigma := \Sigma \cup \{x_0,\dots,x_{m-1}\}$ and
$\widehat\Gamma := \Gamma \cup \{x_0,\dots,x_{m-1}\}$.
Fix an automaton $\calA = \langle Q,\widehat\Gamma,\Delta,q_0,\Omega\rangle$ recognising~$L$.
For $c \in \Sigma$, let $\Pi(c)$~be the set of all profiles of partial runs of~$\calA$
on~$\varphi(\sing(c))$
and, for $i < m$, let $\Pi(x_i)$ be the set of all states from which~$\calA$ accepts
the singleton tree with label~$x_i$.
We construct an automaton~$\calB$ for $\varphi^{-1}[L]$ as follows.
The set of states is $Q \times D \cup \{\langle\bot,\bot\rangle\}$,
where $D$~is the set of priorities used by~$\calA$.
The initial state is $\langle q_0,k\rangle$, for an arbitrary $k \in D$,
and the priority function is given by $\Omega(\langle p,k\rangle) := k$
and $\Omega(\langle\bot,\bot\rangle) = 0$.
The transitions of~$\calB$ are as follows.
\begin{align*}
  &\bigl\langle \langle p,l\rangle,\ c,\
     \langle q_0,k_0\rangle,\dots, \langle q_{n-1},k_{n-1}\rangle\bigr\rangle \quad
  &&\text{for } c \in \widehat\Sigma\,,\ \langle p,k_0,q_0,\dots,k_{n-1},q_{n-1}\rangle \in \Pi(c)\,, \\
  &\bigl\langle \langle \bot,\bot\rangle,\ c,\
     \langle \bot,\bot\rangle,\dots, \langle \bot,\bot\rangle\bigr\rangle
  &&\text{for } c \in \widehat\Sigma\,.
\end{align*}
It is straightforward to check that~$\calB$ accepts a tree $t \in \bbT_m\Sigma$ if, and only if,
$\calA$~accepts~$\varphi(t)$.
\end{proof}

\bibliographystyle{siam}
\bibliography{Regular}

\begin{thebibliography}{10}

\bibitem{AdamekRosicky94}
{\sc J.~{Ad\'amek} and J.~{Rosick\'y}}, {\em {Locally Presentable and
  Accessible Categories}}, Cambridge University Press, 1994.

\bibitem{BlumensathXX}
{\sc A.~Blumensath}, {\em {Branch-Continuous Tree Algebras}}.
\newblock arXiv:1807.04568.

\bibitem{BlumensathYY}
\leavevmode\vrule height 2pt depth -1.6pt width 23pt, {\em {$\omega$-Forest
  Algebras and Temporal Logics}}.
\newblock in preparation.

\bibitem{Blumensath11c}
\leavevmode\vrule height 2pt depth -1.6pt width 23pt, {\em {Recognisability for
  algebras of infinite trees}}, Theoretical Computer Science, 412 (2011),
  pp.~3463--3486.

\bibitem{Blumensath13a}
\leavevmode\vrule height 2pt depth -1.6pt width 23pt, {\em {An Algebraic Proof
  of Rabin's Tree Theorem}}, Theoretical Computer Science, 478 (2013),
  pp.~1--21.

\bibitem{Bojanczyk15}
{\sc M.~Boja{\'n}czyk}, {\em Recognisable languages over monads}.
\newblock unpublished note, arXiv:1502.04898v1.

\bibitem{BojanczykId09}
{\sc M.~{Boja\'nczyk} and T.~Idziaszek}, {\em {Algebra for Infinite Forests
  with an Application to the Temporal Logic EF}}, in {Proc.\ 20th International
  Conference on Concurrency Theory, \smaller{CONCUR}, \smaller{LNCS 5710}},
  2009, pp.~131--145.

\bibitem{BojanczykIdSk13}
{\sc M.~{Boja\'nczyk}, T.~Idziaszek, and M.~Skrzypczak}, {\em {Regular
  languages of thin trees}}, in {Proc.\ 30th International Symposium on
  Theoretical Aspects of Computer Science, \smaller{STACS} 2013}, 2013,
  pp.~562--573.

\bibitem{BojanczykKlin17}
{\sc M.~Boja{\'n}czyk and B.~Klin}, {\em {A non-regular language of infinite
  trees that is recognizable by a sort-wise finite algebra}}.
\newblock arXiv:1804.06667.

\bibitem{Borceux94b}
{\sc F.~Borceux}, {\em {Handbook of Categorical Algebra}}, vol.~2, Cambridge
  University Press, 1994.

\bibitem{MiliusUrbat19}
{\sc S.~Milius and H.~Urbat}, {\em {Equational Axiomatization of Algebras with
  Structure}}, in {Proc.\ 22nd International Conference on Foundations of
  Software Science and Computation Structures, \smaller{FOSSACS} 2019}, 2019,
  pp.~400--417.

\bibitem{Niesser02}
{\sc F.~{Nie\ss er}}, {\em {Non-deterministic Tree Automata}}, in {Automata,
  Logic, and Infinite Games}, E.~Gr\"adel, W.~Thomas, and T.~Wilke, eds.,
  {\smaller{LNCS} 2500}, Springer Verlag, 2002, pp.~149--156.

\bibitem{PerrinPin04}
{\sc D.~Perrin and J.-E. Pin}, {\em {Infinite Words -- Automata, Semigroups,
  Logic and Games}}, Elsevier, 2004.

\bibitem{Reiterman82}
{\sc J.~Reiterman}, {\em {The Birkhoff theorem for finite algebras}}, Algebra
  Universalis, 14 (1982), pp.~1--10.

\bibitem{Schutzenberger:1965il}
{\sc M.~P. Sch{\"u}tzenberger}, {\em {On Finite Monoids Having Only Trivial
  Subgroups}}, Information and Control, 8 (1965), pp.~190--194.

\bibitem{Thomas90}
{\sc W.~Thomas}, {\em {Automata on Infinite Objects}}, in {Handbook of
  Theoretical Computer Science}, J.~van Leeuwen, ed., vol.~B, Elsevier,
  Amsterdam, 1990, pp.~135--191.

\bibitem{Thomas97}
\leavevmode\vrule height 2pt depth -1.6pt width 23pt, {\em {Languages,
  Automata, and Logic}}, in {Handbook of Formal Languages}, G.~Rozenberg and
  A.~Salomaa, eds., vol.~3, Springer, New York, 1997, pp.~389--455.

\bibitem{YanovMuchnik59}
{\sc Y.~I. Yanov and A.~A. Muchnik}, {\em {Existence of $k$-valued closed
  classes without a finite basis}}, Dokl. Akad. Nauk., 127 (1959), pp.~44--46.

\end{thebibliography}

\end{document}